\documentclass{ptptex}

\usepackage{wrapft}
\usepackage{amsmath,amssymb,latexsym,graphicx}

\newcommand{\Slash}[1]{\ooalign{\hfil/\hfil\crcr$#1$}}

\title{Complex Action, Prearrengement for Future and Higgs Broadening}

\author{Holger Bech {\sc Nielsen}$^{1,}$\footnote{On leave of absence to CERN until 31 May, 2008.}
and 
Masao {\sc Ninomiya}$^{2,}$\footnote{Also working at Okayama Institute for Quantum Physics, Kyoyama-cho 1-9, Okayama-city 700-0015, Japan.}}

\inst{
$^1$Niels Bohr Institute, University of Copenhagen,\\%
17 Blegdamsvej Copenhagen \o, Denmark
\\
$^2$Yukawa Institute for Theoretical Physics, Kyoto University, \\
Kyoto 606-8502, Japan

\bigskip
PACS numbers: 12.90.tb, 14.80.cpand, N1.10.-Z}

\abst{We develop some formalism which is very general Feynman path integral in the case of the action which is allowed to be complex. The major point is that the effect of the imaginary part of the action (mainly) is to determine which solution to the equations of motion gets realized. We can therefore look at the model as a unification  of the equations of motion and the ``initial conditions". We have already earlier argued for some features matching well with cosmology coming out of the model.

A Hamiltonian formalism is put forward, but it still has to have an extra factor in the probability of a certain measurement result involving the time after the measurement time.

A special effect to be discussed is a broadening of the width of the Higgs particle. We reach crudely a changed Breit-Wigner formula that is a normalized square root of the originally expected one.}

\begin{document}
\maketitle

\section{Introduction}

We have already in a series of articles~\cite{rf:1,rf:2,rf:3} studied a model in which the initial state of the Universe~\cite{rf:4} is described by a probability density $P$ in phase space, which can and is assumed to depend on what happens along the solution associated {\em at all times} in a formally time translational invariant manner. We shall here repeat and expand on the claim that allowing the action to be complex is rather to be considered as making an assumption less than being a new assumption. In fact we could look at the Feynman path integral: 
\begin{align}
	\int e^{\frac{i}{\hbar}S[path]}Dpath.
\end{align}
Then we notice that whether the action $S[path]$ as usually assumed is real or whether it, as in the present article, should be taken to be complex, the integrand $e^{\frac{i}{\hbar}S[path]}$ of the Feynman-path-way integral is anyway complex. Let us then argue that thinking of the Feynman path integral as the fundamental representation of quantum mechanics it is the integrand $e^{\frac{i}{\hbar}S[path]}$ rather than $S[path]$ itself, which is just its logarithm, that is the most fundamental. In this light it looks rather strange to impose the reality condition that $S[path]$ should be real. If anything one would have though it would be more natural to impose reality on the full integrand $e^{\frac{i}{\hbar}S[path]}$, an idea that of course would not work at all phenomenologically.

But the model that there is no reality restriction on the integrand $e^{\frac{i}{\hbar}S[path]}$ at all and thus also no reality restriction on $S[path]$ could be quite natural and it is -we could say- the goal of the present article to look for implications of such an in a sense simpler model than the usual ``action-being-real-picture". That is to say we shall imagine the action $S[path]$ to be indeed complex 
\begin{align}
	S[path]=S_R[path]+iS_I[path].
\end{align}
The natural -but not strongly grounded- assumption would then be that both the real part $S_R[path]$ and the imaginary part of the action can -for instance in the Standard Model- be written as a four dimensional integrals 
\begin{align}
	& S_R[path]=\int L_R(x)d^4x, \nonumber \\
	& S_I[path]=\int L_I(x)d^4x
\end{align}
where the complex Lagrangian density 
\begin{align}
	L(x)=L_R(x)+iL_I(x)
\end{align}
was split up into the real $L_R$ and imaginary $L_I$ parts each of which is assumed to be of the same form as the usual Standard Model Lagrangian density. However the coefficients to the various terms could be different for real and imaginary part. We could say that the fields, the gauge fields $A_{\mu}^a(x)$, and the fermion fields $\psi_{\alpha}^{(f)}(x)$ and the Higgs field $\phi_{HIGGS}(x)$ obey the same reality conditions as usual (in the Standard Model) so that the action is only made complex by letting the coefficients $\frac{-1}{4g_a^2},\> Z^{(f)},\> m_H^2,\> Z_{HIGGS}$ and $\lambda$ in the Lagrangian density 
\begin{align}
	L(x)
	=\sum_a & \frac{-1}{4g_a^2}F_{\mu \nu}^a(x)F^{a\mu \nu}(x)
	+Z^{(f)}\psi^{(f)}\Slash{D}\psi^{(f)} \nonumber \\
	& +Z_{HIGGS}|D_{\mu}\phi_{HIGGS}(x)|^2-m_H^2|\phi_{HIGGS}(x)|^2
	-\frac{\lambda}{4}|\phi_{HIGGS}(x)|^4
\end{align}
be complex. For instance we imagine the Higgs-mass square coefficient to split up into a real and an imaginary part 
\begin{align}
	m_H^2=m_{HR}^2+im_{HI}^2.
\end{align}

In the following section 2, we shall argue that in the classical approximation as usually extracted from the Feynman path integral it is only the real part of the action $S_R[path]$ that matters. In the following section 3, we then review that the role of the imaginary part $S_I[path]$ is to give the probability density $P\propto e^{-\frac{2}{\hbar}S_I[path]}$ for a certain solution path being realized and we shall explain how the imaginary part $S_I$ takes over the role of the boundary conditions so that we can indeed work with Feynman path integrals corresponding to paths extended through all times rather than just a time interval of interest and essentially ignore further boundary conditions. In such an interpretation it is necessary with a bit of extra assumptions to obtain quantum machanics as we shall review in section 4 and then in secton 5 we develop a Hamiltonian formalism in which the use of the non-hermitian Hamiltonian now is so as to still ensure that rudiment of
  unitarity that says that the collected probability of all the outcomes of a measurement is still as in the usual theory unity. I  section 6 we present assumptions for our interpretation discussed in the precceding sections. Then in section 7 we argue how to derive quantum mechanics under normal conditions. In section 8 we again return to Hamiltonian development. In section 9 we make some discussion of the interpretation of our model. In section 10 we present expected effects when performing the experiment. In section 11 we shall look at the prediction of broadening of the Higgs resonance peak in an interesting way. We then argue in section 12 that the Higgs lifetime may be broadening in our theory. In section 13 we shall draw some conclusions analogous outlook.

\section{Classical approximation only uses $S_R[path]$}

Since the usual theory works well without any imaginary part $S_I$ in the action we must for good phenomenology in first approximation have that this imaginary part is quite hidden. Here we shall now show that as far as the classical approximation to the model is concerned the effect of $S_I[path]$ is indeed negligible and the equations of motion take the almost usual form 
\begin{align}
	\delta S_R=0, 
\end{align}
just it is the {\em real part} $S_R$ rather than the full action $S=S_R+iS_I$ which determines the classical equations of motion.

The argument for the relevance of only the real part $S_R$ in the classical equation of motion is rather simple if one remembers how the classical equations of motion are derived from the path-way-integral in the usual case with its only real action. The argument really runs like this: When we have that $\delta S_R\neq 0$ it means that the real part of the action $S_R$ varies approximately linearly under variation of the path (in the space of all the paths) in a neighbourhood. This, however, means then that the factor $e^{\frac{i}{\hbar}S_R}$ in the Feynman path way integrand oscillate in sign or rather in phase so that -unless the further factor $e^{-\frac{S_I}{\hbar}}$ varies extremely fast- the contributions with the factor $e^{\frac{i}{\hbar}S_R}$ deviating in sign (by just a minus say) will roughly cancel out. So locally we have essentially cancelling out of neighbouring contributions in any neighbourhood where $\delta S_R\neq 0$. We can say that this cancellation gets f
 ormally very perfect in the limit of the coefficient $\frac{1}{\hbar}\to \infty$ in front of $S_R$ in the exponents. This is actually the type of argument used in the usual case of only $S_R$ being present: We can even count on the linear term in the Taylor expansion of 
\begin{align}
	S_R=S_R[path_0]+\Delta path\cdot \frac{\delta S_R}{\delta path}
	+\frac{1}{2}(\Delta path)^2\cdot 
	\frac{\delta^2S_R}{\delta path\> \delta path}+\cdots
\end{align}
dominating the phase rotation for $\hbar$ small, when we look at a region of the order of the phase rotation ``wave length". If indeed also the $S_I$ varies with the same rate the argument strictly speaking breaks down even for $\hbar$ being small. We may, however, argue that in order to find a highly contributing path we shall search in a region not so far from a minimum of $S_I$ and thus $S_I$ will vary relatively slowly - but also we may be interested in $S_R$ near an extremum so this does not really mean that we can count on the rate of variations being so different. We may make the argumentation for that in pracsis the variation of $S_I$ is not so strong compared to the variation of the real part $S_R$ better refering to that we in early articles on our model have argued for that the contributions to $S_I$ from the present cosmological era are especially low compared to the more normal size ones from some very early big bang era. The point indeed were that in the present
  times we are mainly concerned with massless particles -for which the eigentimes are always zero- or non relativistic conserved particles for which the eigentimes are approximately the coordinate time and at the end given just as the universe lifetime. Since the density of particles and thereby the interaction is also today low compared to early cosmological big bang times the contributions today to the imaginary part $S_I$ would mainly come from the passages of the particles from one interaction to the next one and thus like we know for the real part due to Lorentz invariant requirements be proportional to the eigentimes: 
\begin{align}
	S_{I\> contribution}\propto \tau_{eigen}.
\end{align}
Since these eigentimes as we just said are rather trivial, zero or constant, in the today era we expect by far the most impotant variations of the imaginary action $S_I[path]$ to come from variations of the path in the Big Bang times rather than in our times.

Thus essentially when we discuss variations of the path w.r.t. variable variations in our times we expect $\delta S_I$ to be small and the cancellation to occur unless $\delta S_R\big|_{\text{our time}}=0$. So we believe to have good arguments for the classical equations of motion with only use of the real part $S_R$ only to come out even though fundamentally the action would be comlex $S=S_R+iS_I$.

As long as we look for regions in real path-space it is, however, clear that it is the $S_R$ that gives the sign oscillation and thus the cancellation effect wherever then $S_R\> \Slash{\simeq}\> 0$. This in turn means that we only obtain appreciable contributions to the Feynman path integral from the neighbourhoods of paths with the property 
\begin{align}
	\delta S_R=0.
\end{align}
This is thus a derivation of the classical equations of motion as an equation to be obeyed for those paths in the neighbourhood of which an appreciable contribution can arise. Only in the neighbourhoods of the solutions to the classical equations of motion $\delta S_R=0$ do the different noighbouring contributions to the Feynman path integral act in a collaborative manner so that a big result appears.

This result suggesting that it is mainly $S_R$ that determines the classical equation of motion is of course rather crucial for our whole idea, because it means that in the first approximation -the classical one and not overly strong $S_I$- we can hope that it is only $S_R$ that determines the equations of motion. 

\section{Classical meaning of $S_I$}

Even after we have decided that there are such sign oscillation cancellations that all contributions to the Feynman-path integral (and thus assuming Feynman path integrals as the fundamental physics) not obeying the classical equations of motion $\delta S_R[path]=0$ completely cancels out, there are still a huge set of classically allowed paths obeying these equations of motion. The paths in neighbourhoods -in some crude or principal sense of order $\hbar$ expansion- around the classical solutions (to $\delta S_R=0$) are not killed by the cancellations and they have still the possibility for being important for the description by our model. Now each classical solution, say 
\begin{align}
	clsol=some\> path
\end{align}
obeying 
\begin{align}
	\delta S_R[clsol]=0
\end{align}
gives like any other path rise to an $S_I$ value $S_I[clsol]$.

Even without being so specific as we were in last years Bled-proceeding~\cite{rf:1} on this model but just arguing from what everybody will accept about Feynman-path integral interpretation we could say:

Clearly the contribution to the Feynman path integral from a specific classical solution neighbourhood must contain a factor 
\begin{align}
	\int_{\text{\tiny NEIGHBOURHOOD OF clsol ONLY}} 
	e^{\frac{i}{\hbar}S}Dpath \propto 
	e^{-\frac{1}{\hbar}S_I[clsol]}.
\end{align}
Since we all accept a loose statement like ``the probability is given by numerically squaring the Feynman path integral (contribution)" we may accept as almost unavoidable -whatever the exact interpretation scheme assumed- that the probability for the classical solution $clsol$ being (the?) realized one must be proportional to 
\begin{align}
	P[clsol]\propto \left|e^{-\frac{1}{\hbar}S_I[clsol]}\right|^2
	=e^{-\frac{2}{\hbar}S_I[clsol]}.
\end{align}
This probability density over phase space of initial conditions $P[clsol]$ were exactly what we called $P$ also in the earlier works on our model, where we sought to be more general by not talking about $P[clsol]$ being $e^{-\frac{2}{\hbar}S_I[clsol]}$ but just talking about it as a general probability weight the behavior of which could then be discussed separately.

Let us stress actually that if you do not say anything about the functional behavior of the probability then the formalism with $P$ in our earlier works is so general that it can hardly even be wrong. Of course if you write it as $P[clsol]=e^{-\frac{2}{\hbar}S_I[clsol]}$ and do not assume anything about $S_I$ it remains so general that it hardly can be wrong, because we have just defined 
\begin{align}
	S_I[clsol]=-\hbar\cdot \frac{1}{2}\log \left(P[clsol]\right).
\end{align}
However, if we begin to assume that in analogy to the real part of the action $S_R$ also the imaginary is an integral over time 
\begin{align}
	S_I[path]=\int L_I(t;path)dt
\end{align}
of some Lagrangian $L_I$ in a time translational invariant way or the even more specific form as a space time integral, then we do make nontrivial assumptions about $P=e^{-\frac{2}{\hbar}S_I}$. Usually we would say that we already from well-known (physical) experience, further formalized in the second low of thermodynamics, know that the $S_I[clsol]$ is {\em only allowed} to depend on what goes on along the path $clsol$ at the initial moment of time $t=t_{initial}$. This ``initial time" is imagined to be the time of the Big Bang singularity -if such a singularity indeed existed-. If there were no such initial time (as we suggested in one of the papers in the series on our imaginary action) then one might in the usual theory not really know what to do. Perhaps one can use the Hartle-Hawking no boundary model~\cite{rf:4}, but that would effectively look much like a Big Bang start.

But our present article motivating arguments are: 
\begin{enumerate}
\item An imaginary action is an almost milder assumption than assuming it to be zero $S_I=0$.
\item To assume that the essential logarithm of the probability $P$ namely $S_I$ should depend only on what goes on at a very special moment of time $t=t_{initial}$ sounds almost time non-translational invariant. (Here Hartle-Hawking no boundary may escape elegantly though.)
\end{enumerate}

\subsection{The classical picture in our model resumed}

Let us slightly summarize and put in perspective our classical approximation for our imaginary action model: 
\begin{enumerate}
\item We argued for the classical equations of motion be given alone by the real part of the action $\delta S_R=0$ so that the imaginary part $S_I$ were not relevant at all, so that it were in first approximation not so serious classically whether you assume that $S_I$ is there or not.
\item We argued that the main role of the imaginary part $S_I[clsol]$ of the action were to give a probability distribution over the ``phase space" (it has a natural symplectic structure and is if restricted to a certain time $t=t_0$ simply the phase space) of the set of classical solutions: 
\begin{align}
	P[clsol]=\text{``normalization"}\cdot e^{-\frac{2}{\hbar}S_I[clsol]}.
\end{align}
Since $\hbar$ is small this formula for the density presumably very strongly can derive the ``true" solution to almost the one with the minimal -in the sense of the most negative- $S_I[clsol]$. (But really huge amounts of classical solutions with bigger $S_I$ could statistically take over.)
\item We argued that in the present era -long after Big Bang- the effects of $S_I$ were at least somewhat suppressed due to that now we mainly have non-relativistic conserved particles or massless particles and not much interactions compared to early big bang times.
\end{enumerate}

At this classical stage of the development of our imaginary action component model it will seem to cause lots of prearrangements of events that could cause especially low (i.e. negative) contributions to $S_I$ because the classical solution realized will be one with excepionally presumably numerically large negative $S_I$ so as to make $P\propto e^{-\frac{2}{\hbar}S_I}$ large. Really we could say that it is as if the universe were governed by a leader seeking as his goal to make the imaginary part $S_I$ minimal.

\subsection{Is it possible that we did not discover these prearrangement?}

One reason -and that is an important one- is that the processes in our era involves mainly conserved non-relativistic particlesor totally massless ones (photons) so that the eigentimes which give rise to $S_I$-contributions become rather trivial. But if really that were all then this leader of the development of the universe would make great efforts to either prevent or favour strongly the various relativistic particles accelerators. But one could wonder how we could have discovered whether a certain type of accelerator were disfavoured, because it would very difficult to know how many of them should have been built if there were no $S_I$-effects. Such decisions as to what accelerators to build happens as a function of essentialy a series of logical -and thus presumably given by the equations of motion $\delta S_R=0$- arguments. But then there will be no clear sign that anything were disfavoured or favoured. It might be very interesting to look for if there would be any effec
 t of ``influence from the future" if one let the running or building of some relativistic particle depend on a card-play or a quantum random number generator. If it were say disfavoured by leading to a positive $S_I$-contribution to run an accelerator of the type in question, then the cards would be prearranged so that the card pulled would mean that one should not run the accelerator. By the same decision ``don't run" being given by the cards statistically too many times one might discover such an $S_I$-effect.

It could be discovered in principle also us notice surprisingly bad luck for accelerators of the disfavoured type. But it is not easy because the unlucky accidents could go very far back in time: a race or a culture society long in the past that would have had better chance tallents or interest for building relativistic high energy accelerators could have gone extinct. But it would be hard for us to evaluate which extinct societies in the past had the better potentiality for making high energy accelerators later on. So it could be difficult to notice such $S_I$-effects even if they manage to keep a certain type collision down both in experimental apparatuses and in the cosmic ray.

Only if the bad luck for an accelerator were so lately induced as seemingly were the case with the S.S.C.~\cite{rf:5} collider in Texas, which would have been larger than L.H.C. but which were stopped after one quarter of the tunnel had been built. This were a case of so remarkably bad luck that we may (almost) take as an evidence for some $S_I$-effect like effect and that some of the particles to be produced -say Higgses- or destroyed -say baryon-number- made up something unwanted when one seeks to minimize $S_I$.

\subsection{Do we expect card game experiments to give results?}

Before going to quantum mechanics let us a moment estimate how much is needed for a card game or quantum number generator decission on say the switching on of a relativistic accelerator could be expected to influence backwards in time~\cite{rf:8} the a priori random number (the card pull or the quantum random number) generated:

The imaginary action will in both cases accelerator switched on or not switched on get possibly much bigger contributions from the future. These future contributions are from our point of view extremely difficult to calculate, alone e.g. the complicated psychological and political consequences of a certain run of the accelerator on if and how much it will be switched on later would be exceedingly difficult to estimate. So in pracsis we must suppose that after a certain card game determined switch on or off there will come a future witha in pracsis random $S_I$-contribution $S_{I\> future}$ depending on the switch on or off in a {\em random way}. So unless for some reason the contribution from the switch on or switch off time is bigger than or comparable to the fluctuations with the switch on or off $\Delta_{on/off}S_{I\> future}$ i.e. unless 
\begin{align}
	S_I\Big|_{\text{accelerator contribution}}\gtrsim 
	\Delta_{on/off}S_{I\> future}
	\label{ln60}
\end{align}
we will not see any effects of $S_I$ in such an experiment. Now a very crude first orientation consists in estimating that the space-time region over which the switch on or off can influence the future is the whole forward light cone starting from the accelerator decission site.

Even if the sensitivity of $S_I$ from most of the consequences the on/off decission may have by accident in this light cone is appreciably lower than the sensitivity to the particles in the accelerator, there is a huge factor in space-time volume to compete against in order that (\ref{ln60}) shall be fulfilled. This is a big factor even if we take into account that the light cone space-time volume is random so that it is the square 
\begin{align}
	\left(\Delta_{on/off}S_{I\> future}\right)^2\propto 
	\text{\em Vol }(\text{light cone})
\end{align}
that is proportional to the forward light cone space-time volume rather than the fluctuation itself 
\begin{align}
	\Delta_{on/off}S_{I\> future}\propto 
	\sqrt{\text{\em Vol }(\text{light cone})}
\end{align}
going rather like the square root.

\subsection{Hypothetical case of no future influence}

If, however, we were thinking of the very unusual case that two different random number decissions had no difference in their future consequences at all, then of course we would have no fluctuations in $S_I$ from the future and thus $\Delta_{on/off}S_{I\> future}=0$. In such an unrealistic case of all tracks of the decission being immediately totally hidden there is no way that in our model then the effect from the accelerator on or off time could be drowned in the future contribution fluctuations.

\section{Quantum effects}

Really the at the end of last section mentioned special case of a decission being quantum random say but being forever hidden so that it cannot influence the future and thereby the future contribution $S_{I\> future}=\int_{now}^{\infty}L_Idt$, is the one you have in typical quantum mechanical experiments. In for instance a typical quantum experiment one starts by preparing a certain unstable particle and then later measure the energy of the decay products from the decay.

We could in this experiment look at the actual life time of the unstable particle $t_{actual}$ as a quantum random number -a quantum random number decission of the actual life time of just the particle in question-. But if one now measures the energy of the decay products -the conjugate variable to the actual time $t_{actual}$- it is impossible that the actual time $t_{actual}$ shall ever been known. So here we have precisely a case of a decission which is kept absolutely secret. But that then means that the future cannot know anything about the actual life time $t_{actual}$ and $S_{I\> future}$ can have no $t_{actual}$ dependence. Thus in this case the fluctuation 
\begin{align}
	\Delta_{t_{actual}}S_{I\> future}=0
\end{align}
of $S_{I\> future}$ due to the variation of $t_{actual}$ must be zero. Thus in this case of such a hidden decission there is no way to get the $S_I$ contribution from the existence time of the unstable particle $S_I\big|_{\text{from $t_{actual}$}}$, which is presumably proportional to $t_{actual}$ 
\begin{align}
	S_{I}\big|_{\text{from $t_{actual}$}}=\frac{\Gamma_I}{2}t_{actual}, 
\end{align}
dominated out by the future contribution $S_{I\> future}$. So if truly in some sense the coefficient here called $\frac{\Gamma_I}{2}$ giving the $S_I$-contribution $S_I\big|_{\text{from $t_{actual}$}}$ is large because of being inversely proportional to $\hbar$, then there should be strong effects of $S_I$ in this case, or rather effects that cannot be excused as being just accidental. Really the philosophy of our model which we are driving to as that the effects of $S_I$ are indeed huge but they come in by prearrengement so that whatever happens comes seemingly for us the likely and natural consequences of what already happened ealier. Thus the fact that certain particles or certain happenings are getting indeed strongly prevented by such prearrengements is not noticed by us.

In the case of an actual decay time $t_{actual}$ which similarly to the slit passed in the by Bohr and Einstein discussed double slit experiment does not have any correlation neither with prior to experiment nor to later than experiment times there is nothing that can overwrite/dominate the effect out.

So we say that in such a never measure but by quantum random number way chosen variable as $t_{actual}$ the $S_I$-effect shoud show up. But now of course there is a priori the difficulty that if precisely the actual lifetime $t_{actual}$ is \underline{not} measured, then how do we know if it were systematicly made shorter in our model than in the real action model? Well since we do not measure it -if we did we would make the effect be overshadowed by accidental effects from future- we cannot plot its distribution and check that it is stronger peaked towards zero than the theoretical decay rate calculation would say it should be. We can, however, use Heisenberg uncertainty principle and should in our model find that Breit-Wigner distribution of the decay product energy (=invariant mass) has been broadened compared to a real action theory. Since we shall suggest that it is likely that especially the Higgs particle will show very big $S_I$-effects it is especially the Higgs Brei
 t-Wigner we suspect to be significantly broadened.
\subsection{Quantum experiment formulation}

The typical quantum experiment which we should seek to describe in our model is of the type that one prepares some state $|i\rangle$ -in the just discussed case an unstable particle, a Higgs e.g.- and then measure an outcome $|f\rangle$, which in the case we suggested would be the decay products -$b\bar{b}$ jets say- with a given energy or better invariant mass. When one has prepared a state $|i\rangle$ it means that one is scientifically sure that one got just that for the subsystem of the universe considered. Thus whether to reach that state were very suppressed or favoured by the $S_I$-effects does no longer matter because we know we got it ($|i\rangle$) already. We should therefore so to speak normalize the chance for having gotten $|i\rangle$ to be zero even if this would not be one would have theoretically calculated in our model. One should have in mind that since our model is in principle also a model for the realized solution or the initial state conditions one could
  ideal by calculate the probability that at the moment of time of the start of the experiment, say $t_i$, the Universe is indeed in the state $|i\rangle$ (or that the subsystem of the Universe relevant for the experiment is in a state $|i\rangle$). In pracsis of course such calculations are not possible -except perhaps and even that is optimistic some cosmological questions as the Hubble expansion of the energy density in the universe-.

\subsection{Practical quantum calculation, ignoring outside regions in time}

In the typical quantum experiment -as already alluded to- we have the system first in a state $|i\rangle$ at $t=t_i$ say and then later at $t=t_f$ observe it in $|f\rangle$.

Then one would using usual (meaning real action) Feynman path integral formulation say that the time development transition amplitude from $|i\rangle$ at $t_i$ to $|f\rangle$ at $t_f$ is given as 
\begin{align}
	\langle f|U|i\rangle =
	\int e^{\frac{i}{\hbar}S[path]}\langle f|path(t_f)\rangle 
	\langle path(t_i)|i\rangle Dpath
	\label{ln85}
\end{align}
where $\langle path(t_i)|i\rangle$ is the wave function of the state $|i\rangle$ expressed in terms of the field configuration value $path(t_i)$ of the path $path$ taken at time $t_i$ and $\langle f|path(t_f)\rangle$ in the same way is the wave function for the state $|f\rangle$ expressed by the value of the path $path$ at time $t_f$. The paths integrated functionally given in (\ref{ln85}) are in fact only paths describing a thinkable time development in the time interval $[t_i,t_f]$.

We can easily say that in our model we now insert our complex $S[path]$ instead of the purely real one in the usual theory. But that is not in principle the full story in our model for a couple of reasons:

If we constructed from (\ref{ln85}) all the amplitudes obtained by inserting a complete set of $|f\rangle$ states, say $|f_j\rangle ,\> j=1,2,\cdots$, with $\langle f_j|f_k\rangle =\delta_{jk}$ instead of $|f\rangle$ and then summed the numerical squares 
\begin{align}
	\sum_{j}\left|\langle f_j|U|i\rangle \right|^2=
	\left\{ \begin{array}{ll}
	1, & \text{in usual theory} \\
	\text{not }1, & \text{in our theory usually}
	\end{array}\right.
\end{align}
we would not in our model get unity in our model such as one gets in the usual real action theory. This is of course one of the consequences of that our development matrix $U$ (essentially S-matrix) is not unitary.

However, we have in our model taken a rather timeless perspective and we especially take it as given from the outset that the world exists at all times $t$. So we cannot accept that the probability for the universe existing at a later time should be anything else than unity. So we must take the point of view that when we have seen that we truly got $|i\rangle$ then the development matrix $U$ (essentially the S-matrix) can only tell us about the relative probability for the various final state $|f\rangle$ we might ask about, but the probabilities for a complete set must be normalized to unity. This argumentation at first suggest the usual expression $\left|\langle f|U|i\rangle \right|^2$ to be normalized to 
\begin{align}
	P(|f\rangle \big||i\rangle )=
	\frac{\left|\langle f_j|U|i\rangle \right|^2}{||U|i\rangle ||^2}.
	\label{ln92}
\end{align}
so that we ensure 
\begin{align}
	\sum_j P(|f_j\rangle \big||i\rangle )=1.
\end{align}
However, this expression is not exactly -although presumably a good approximation- to the prediction of our model. The point is that we have in our model even influence from the future contribution to $S_I$. Typically we already suggested that these contributions $S_{I\> future}$ would vary strongly -but not in most cases so that we have any way to know how- and so we really expect that one of the possible measurement results $|f_j\rangle$ will be indeed favoured strongly by giving rise to the most negative $S_{I\> future}\big|_{|f_j\rangle}$. Since we, however, do not know how to calculate which $|f_j\rangle ,\> j=1,2,\cdots$, gives the minimal $S_{I\> future}\big|_{|f_j\rangle}$. We in pracsis would make the statistical model of putting this factor $\exp\left\{-2S_{I\> future}\big|_{|f_j\rangle}\right\}$ in the probability 
\begin{align}
	P(|f_j\rangle \big||i\rangle )\stackrel{\text{our model}}{=}
	\frac{\left|\langle f_j|U|i\rangle \right|^2}{||U|i\rangle ||^2}
	\cdot \exp \left\{-2S_{I\> future}\big|_{|f_j\rangle}\right\}
	\label{ln95}
\end{align}
equal to a constant $1$. Only in the case we would decide to use the result $j$ of the measurement to e.g. decide whether to start or not start some very high $S_I$-procucing accelerator as presumably S.S.C would have been would we expect that we should use (\ref{ln95}) rather than simply (\ref{ln92}). But already (\ref{ln92}) is interesting and unusual because it for instance contains the Higgs broadening effect, which we suggest that one should look for at L.H.C. and the Tevatron~\cite{rf:6}. We shall go this in the later sections.

\section{Quantum Hamiltonian formalism}

Let us, however, first remind a bit about last years Bled talk on this subject and give a crude idea about one might write Feynman diagrams for evaluation of our expression (\ref{ln92}).

First it is rather easy to see that the usual (i.e. with real action) way of deriving the Hamiltonian development in time takes over practically just by saying that now the coefficients in the Lagrangian or Lagrangian density are to be considered complex rather than just real. The transition from Feynman-path integral to a wave function and Hamiltonian description is, however, whether the Lagrangian is real or not connected with constructing a measure $D$ in the space of field or variable values ata a given time.

Of course the Hamiltonian $H$ derived from the complex action organized to obey say 
\begin{align}
	\frac{d}{dt}U(t_f,t_i)=iH
\end{align}
will not be Hermitean. That is of course exactly what is connected with $U$ not being unitary.

When talking about the wave function and Hamiltonian formulation we have presumably the duty to bring up that according to last years proceedings we take a slightly unusual point of view w.r.t. how we apply the Feynman path way integral. Usually one namely only use the Feynman path integral as a mathematical technique for solving the Schr\"{o}dinger equation. We use, however, in our model as discussed last year the Feynman path integral as the fundamental presentation of the model, Hamiltonian or other formulations should be derived from our a little bit unusual definition of the theory in terms of the Feynman path integral(s).

\subsection{Our ``fundamental" interpretation}

Our slightly modified interpretation of the Feynman path integral is based on the already stressed observation that the imaginary part $S_I$ chooses the initial state conditions or the actually to be realized solution to the equations of motion. This namely, then means that normal boundary conditions become essentially unimportant and that it is thus most elegant to sum over all possible boundary conditions, so that the imaginary part $S_I$ so to speak can be totally free to choose effectively the boundary conditions it would like. Even if one puts in some boundary conditions by hand there only has to be a quite moderate wave function overlap with the initial condition which ``$S_I$ prefers" and that will be the one given the dominant weight even if the moderate overlap is quite small. The $S_I$ in fact goes in the exponent with the big number $\frac{1}{\hbar}$ as a factor and might easily blow a small overlap up to a big part of the Feynman path integral.

We proposed therefore as our outset in the last years proceedings that the probability for the path at some time $t$ passing through a certain range of variables $I$ so that 
\begin{align}
	path(t)\in I
\end{align}
should be given by 
\begin{align}
	P(path(t)\in I)=
	\frac{\displaystyle \sum_{\text{BOUNDARIES}}
	\left|\int e^{\frac{i}{\hbar}S[path]}\chi [path]
	Dpath\right|^2}{\displaystyle \sum_{\text{BOUNDARIES}}
	\left|\int e^{\frac{i}{\hbar}S[path]}Dpath\right|^2}
	\label{ln109}
\end{align}
where the projection functional 
\begin{align}
	\chi [path]=
	\left\{ \begin{array}{l}
	1 \quad \text{for }path(t)\in I \\
	0 \quad \text{for }path(t)\not\in I \\
	\end{array}\right. .
\end{align}
Here of course $path(t)$ stands for the set of values for the fields (or variables $q_k$ in the case of a general analytical mechanical system) at the time $t$ on the path $path$. The ``BOUNDARIES" summed over stands for the boundaries at $t\to -\infty$ and $t\to \infty$ or whatever the boundaries of time may be.

The special point of our model is that the BOUNDARIES are in first approximation not relevant because $S_I$ takes over. The details of how to sum over them is thus also not important. A part of last years formalism were to write the whole functional integral used in (\ref{ln109}) as an inner product of one factor $|A(t)\rangle$ from the past of some time $t$ and one factor $|B(t)\rangle$ from the future of time $t$: 
\begin{align}
	\langle B(t)|A(t)\rangle =\int e^{\frac{i}{\hbar}S[path]}
	Dpath.
\end{align}
We have then defined the two Hilbert space vectors (describing the whole Universe) by means of path integrals over path's running respectively over path's from the beginnings of times (say time $t\to -\infty$) up to the considered time $t$
\begin{align}
	\langle q|A(t)\rangle 
	=\int_{\text{FOR $path$ ON $[-\infty ,t]$ ENDING WITH $path(t)=q$}}
	e^{\frac{i}{\hbar}S_{-\infty ,t}[path]}Dpath
\end{align}
and over paths from $t$ to the ``ends of times" (say $t\to \infty$) 
\begin{align}
	\langle B(t)|q\rangle 
	=\int_{\text{OVER $path$ ON $[t,+\infty ]$ BEGINNING WITH $path(t)=q$}}
	e^{\frac{i}{\hbar}S_{t,+\infty}[path]}Dpath.
\end{align}
Here of course 
\begin{align}
	S_{-\infty ,t}[path]=\int_{-\infty}^tL(path(\tilde{t}))d\tilde{t}
\end{align}
and
\begin{align}
	S_{t,+\infty}[path]=\int_t^{+\infty}L(path(\tilde{t}))d\tilde{t}.
\end{align}
In order that these Hilbert space vectors be welldefined one would usually have to specify the boundary conditions at the beginnings and ends of times, $-\infty$ and $+\infty$, but because of the imaginary part $S_I$ assumed in the present work it will be so that it will be extremely difficult to change the results for $|A(t)\rangle$ and $|B(t)\rangle$ by more than over all factors by modifying the boundary conditions at $-\infty$ and $+\infty$ respectively. In this sense we can say that the Hilbert-vectors $|A(t)\rangle$ and $\langle B(t)|$ are approximately defined without specifying the boundary conditions. Remember it were the main idea that $S_I$ takes over the role of boundary conditions i.e. $S_I$ rather than the boundary conditions choose the initial state conditions. With such a philosophy of $S_I$ fixing the initial state conditions we might be tempted to interprete $|A(t)\rangle$ as the wave function of the whole Universe at time $t$ derived from a calculation usin
 g the initial state conditions given somehow by $S_I$. More interesting than an in pracsis unaccessible wave function $|A(t)\rangle$ for the whole Universe would be a wave function for a part of the universe -say a few particles in the laboratory- and then we might imagine something like that when we have prepared a state $|\psi (t)\rangle$ at time $t$ for some such subsystem of the Universe it should correspond to the state vector $|A(t)\rangle$ factorizing like 
\begin{align}
	|A(t)\rangle =|\psi (t)\rangle \otimes |rest\> A(t)\rangle .
	\label{ln118}
\end{align}
However, this will in general \underline{not} be quite true. Rather we must usually admit that whether we get a welldefined state $|\psi (t)\rangle$ for the particles in the laboratory also will come to depend on $|B(t)\rangle$ and not only on $|A(t)\rangle$.

It is true that in order that our model shall not be immediately killed the $S_I$-dependence in some era prior to our own -presumably the Big Bang times- were much more significant in choosing the right classical solution (and then thereby also approximately the to be realized quantum initial state too) than the present and future eras. Thus in this approximation $|A(t)\rangle$ represents the development from the by the Big Bang times $S_I$-contributions (supposed to be dominant) selected initial state untill time $t$. But although in this first approximation gives that $|A(t)\rangle$ should represent the whole development there are at least some observations that must depend strongly on $|B(t)\rangle$ also. This is the random results which after usual quantum mechanics -``measurements theory"- comes out only statistically predicted. If the $|A(t)\rangle$ state develops into a state in say equal probability of two eigenvalues for some dynamical variable that $|A(t)\rangle$ ca
 n tell us which of the two values in realized. It can, however, in our formalism still depend on $|B(t)\rangle$.

\section{Our interpretation assumption(s)}

In order to see how $|B(t)\rangle$ comes in we have from last year our interpretation assumptions quantum mechanically: 

Let us express the interpretation of our model by giving the expression for the probability for obtaining a set of dynamical variables to at a certain time $t$ have the values inside a certain range $I$ (a certain interval $I$). The answer to each question is what one would usually identify with the expectation value of the projection operator $\mathcal{P}$ projecting on the space spanned by the eigenspaces (in the Hilbert-space) corresponding to the eigenvalues in the range $I$. In usual theory you would write the probability for finding the state $|\psi (t)\rangle$ to give the dynamical variables in the range $I$ would be 
\begin{align}
	P(I)=\frac{\langle \psi (t)|\mathcal{P}|\psi (t)\rangle}
	{\langle \psi (t)|\psi (t)\rangle}
	\label{ln125}
\end{align}
where the denominator $\langle \psi (t)|\psi (t)\rangle$ is not needed if $|\psi (t)\rangle$ already normalized.

But now supposed we also knew about some measurements being done later than the time $t$. Let us for simplicity imagine that one for some simple system -a particle- managed to measure a complete set of variables for it. Then one would know a quantum state in which this particle did end up. Say we call it $|\phi_{END}\rangle$. Then we would be tempted to say that now -with this end up knowledge- the probability for the particle having at time $t$ its dynamical variables in the range $I$ would be 
\begin{align}
	P(I)=\frac{\left|\langle \phi_{END}(t)|\mathcal{P}|\psi (t)\right|^2}
	{\langle \phi_{END}(t)|\phi_{END}(t)\rangle 
	\langle \psi (t)|\psi (t)\rangle}
	\label{ln127}
\end{align}
where $|\phi_{END}\rangle (t)$ means the state develpped back (in time) to time $t$.

But the question for which we here wrote a suggestive answer were presumably not a good question because: One would usually require that if we ask for whether the variables are in the interval $I$ then one should measure if they are there or not. Such a measurement will, however, typically interfere with the particle so that later extrapolate back the end state -if one could at all find it- to time $t$ i.e. find $|\phi_{END}\rangle (t)$ sounds impossible.

You might of course ignore the requirement of really measuring if the system (particle) at time $t$ is interval $I$ and just say that (\ref{ln127}) could be true anyway; but then it is not of much value to know $P(I)$ from (\ref{ln127}) if it is indeed untestable in the situation. You might though ask if expression (\ref{ln127}) could at least be taken to be true in the cases where it \underline{were} tested. There are some obvious consistency checks connected it to measurable questions: You could at least sumover a complete set of $|\phi_{END}\rangle$ states and check that get the measurable (\ref{ln125}) back.

The from the measurement point of view not so meaningfull formula (\ref{ln127}) has in its Feynman integral form we could claim a little more beauty than the more meaningfull (\ref{ln125}) because we in (\ref{ln127}) can say that we stick in the projection operator $\mathcal{P}$ just at the moment of time $t$, but basically use the full Feynman path integral otherwise from the starting to the final time: 
\begin{align}
	P(I)\Big|_{\text{from (\ref{ln127})}}=
	\frac{\left|\langle \phi_{END}|\int 
	e^{\frac{i}{\hbar}S_{t_s,t_f}[path]}\mathcal{P}
	\Big|_{\text{insert at $t$}}Dpath|\psi \rangle\right|^2}
	{\left|\langle \phi_{END}|\int 
	e^{\frac{i}{\hbar}S_{t_s,t_f}[path]}Dpath
	|\psi \rangle\right|^2}.
	\label{ln131}
\end{align}
Here the paths are meant to be paths defined on the time interval $t_s$ where one get the starting state $|\psi \rangle =|\psi (t_s)\rangle$ to the end time at which one sees the final state for the system $|\phi_{END}\rangle$. One then uses in formulae (\ref{ln131}) both the Feynman path integral to solve the Schr\"{o}dinger equation to develop $|\psi (t_s)\rangle$ forward and $|\phi_{END}$ backward in time.

The reason that we discuss such difficult to associate with experiment formulas as (\ref{ln127}) and the equivalent (\ref{ln131}) is that it is this type of expression we postulated to be the starting interpretation of our model. In fact we postulated (\ref{ln131}) but without putting any boundary conditions $|\psi (t_s)$ and $|\phi_{END}$ in and letting $t_s\to -\infty$ and $t_f\to +\infty$. It is of course natural in our model to avoid putting in boundary conditions, since as we have told repeatedly the imaginary action does the job instead. Without these boundary conditions specified by $|\psi \rangle$ and $|\phi_{END}$ or with boundary conditions summed -as will make only little difference once we have $S_I$- the formulas come to look even more elegant: Our model is postulated to predict for the probability for the variables at time $t$ passing the range $I$ to be 
\begin{align}
	P(I)=
	\frac{\displaystyle \sum_{\text{BOUNDARIES}}\left|
	\int e^{\frac{i}{\hbar}S[path]}\mathcal{P}\Big|_{\text{at $t$}}
	Dpath\right|^2}{\displaystyle \sum_{BOUNDARIES}\left|\int 
	e^{\frac{i}{\hbar}S[path]}Dpath\right|^2}.
	\label{ln134}
\end{align}
As just said the summing over the boundary states BOUNDARIES is expanded to be only of very little significance in as far as $S_I$ should make some boundaries so much dominate that as soon as a bit of the dominant one is present in a random boundary it shall take over.

The expression (\ref{ln134}) is practically the only sensible proposal for interpreting a model in which the Feynman path integral is postulated to be the fundamental physics. If we for instance think of $I$ as a range of dynamical variables which are of the types used in describing the paths, then what else could we do to find the contribution -to the probability or to whatever- than chopping out those paths which at time $t$ have $path(t)\in I$. But such a selection of those paths going at time $t$ through the interval $I$ corresponds of course exactly to inserting at time $t$ the projection operator $\mathcal{P}$ corresponding to a subset of the variables used to describe the paths. In quantum mechanics one always have to numerically square the ``amplitude" which is what you get at first from the Feynman path way integral, so there is really not much possibility for other interpretation than ours once has settled on extracting the interpretation out of a Feynman path integ
 ral with paths describing thinkable developments in configuration (say $q$) space through \underline{all} times.

Once having settled on such an interpretation for the configuration space variables -supposed here used in the Feynman path description- by formula (\ref{ln134}) and having in mind that at least crudely $|A(t)\rangle$ is the wave function of the Universe it is hard to see that we could make any other transition to the postulate of the probability for an interval $I$ involving also conjugate momenta than simply to put in the projection operator $\mathfrak{P}$ anyway.

The formula for the probability of passage of the range $I$, formula (\ref{ln134}), for which we have argued now that it is the only sensible and natural one to get from Feynman path integral using all times is in terms of our $|A(t)\rangle$ and $|B(t)\rangle$ written as 
\begin{align}
	P(I)=
	\frac{\displaystyle \sum_{\text{BOUNDARIES}}\left|\langle B(t)|
	\mathcal{P}|A(t)\rangle\right|^2}
	{\displaystyle \sum_{BOUNDARIES}\left|\langle B(t)
	|A(t)\rangle \right|^2}.
	\label{ln140}
\end{align}

\section{How to derive quantum mechanics under normal conditions}

This formula (\ref{ln140}) although nice from the estetics of our Feynman-path-way based model is terribly complicated if you would use it straight away.

In order that our model should have a chance to be phenomenological viable it is absolutely needed that we can suggest a good approximation (scheme), in which it leads to usual quantum mechanics with its measurement-``theory", with the usual only statistical predictions.

It is easily seen from (\ref{ln140}) that what we really need to obtain the usual -and measurementwise meaningfull- expression (\ref{ln125}) is the approximation 
\begin{align}
	|B(t)\rangle \langle B(t)|\propto {\bf 1}
	\label{ln142}
\end{align}
where {\bf $1$} is the unit operator in the Hilbert space.

\subsection{The argument for the usual statistics in quantum mechanics}

This approximation (\ref{ln142}) is, however, not so difficult to give a good argument for from the following assumption which are quite expected to be true in pracsis in our model: 
\begin{enumerate}
\item Although the $S_I$-variations that gives rise to selection of the to be realized solutions to the classical equations of motion are supposed to be much smaller in future (i.e. later than $t$) times than on the past side of $t$ they are the only ones that can take over the boundary effects on the future side and thus determine $|B(t)\rangle$. However, especially since they are relatively weak $S_{I\> future}$ terms it may be needed to look for enormously far futures to find the contributions at all. 

Thus one has to integrate the equations of motion $\delta S_R=0$ over enormously long times to get back from the ``future" to time $t$ with the knowledge of the state $|B(t)\rangle$ which is the one favoured from the $S_I$-contributions of the future.
\item Next we assume that the equations of motion in this future era are effectively sufficiently \underline{ergodic} that under the huge time spans over which they are to be integrated up the point at $t$ in phase space corresponding to the by the future $S_I$-contributions become approximately randomly distributed over the in pracsis useful phase space.
\end{enumerate}
From these assumptions we then want to say that in classical approximation $|B(t)\rangle$ will be a wave packet for any point in the phase space with a phase space constant probability density. That is how a snapshot of an ergodicly developping model looks at a random time after or before the one its state were fixed. When we take the weighted probabilities of all the possible values of $|B(t)\rangle \langle B(t)|$ we end up from this ergodicity argument that the density matrix to insert to replace $|B(t)\rangle \langle B(t)|$ is indeed proportional to the unit operator. I.e. we get indeed from the ``ergodicity" the approximation (\ref{ln142}).

If we insert (\ref{ln142}) into our postulated formula (\ref{ln140}) we do indeed obtain 
\begin{align}
	P(I)=\frac{\langle A(t)|\mathcal{P}|A(t)\rangle}
	{\langle A(t)|A(t)\rangle}
\end{align}
which is (\ref{ln125}) but with $|A(t)\rangle$ inserted for the initial wave function. Hereby we could claim to have derived from assumptions or approximations the usual quantum mechanics probability interpretation.

That we can get this correspondance is of course crucial for the viability of our model.

Let us remark that since $|B(t)\rangle \langle B(t)|\propto {\bf 1}$ is only an approximation the probability $P(I)$ for the interval $I$ being passed at time $t$ depends in principle via $|B(t)\rangle$ on the future potential events to be avoided or favoured.

The argumentation for $|B(t)\rangle \langle B(t)|$ being effectively proportional to unity by the ``ergodicity" is the same as the classical that even if there are some adjustments for future they look for as random except in very special cases.

\section{Returning to Hamiltonian development now of $|A(t)\rangle$}

It is obvious that one can use the non-hermitean Hamiltonian $H$ derived formally from the complex Lagrangian $L=L_R+iL_I$ associated with our complex action $S=S_R+iS_I$ to give the time development of $|A(t)\rangle$ 
\begin{align}
	i\frac{d}{dt}|A(t)\rangle =H|A(t)\rangle .
\end{align}
The analogous development (Schr\"{o}dinger) equation for $|B(t)\rangle$ only deviates by a sign in the time and the Hermitean conjugation in going from ket to bra 
\begin{align}
	i\frac{d}{dt}|B(t)\rangle =H^{\dagger}|B(t)\rangle
\end{align}
so that 
\begin{align}
	i\frac{d}{dt}\langle B(t)|=-\langle B(t)|H
\end{align}
and we thus can get 
\begin{align}
	\frac{d}{dt}\langle B(t)|A(t)\rangle =0.
\end{align}

\subsection{S-matrix-like expressions}

Realistic S-matrix scattering only going on in a small part of the Universe so that one should really imagine $|A(t)\rangle$ factorized into a Cartesian product like (\ref{ln118}), but for simplicity let us (first) take this splitting out of our presentation. That means that we simply assume that by some scientific argumentation has come to the conclusion that one knows the $|A(t)\rangle$ state vector at the initial time $t_i$ for the experiment to be 
\begin{align}
	|A(t_i)\rangle =|i\rangle .
	\label{ln155a}
\end{align}
Then the looking for the final state $|f\rangle$ at the somewhat later time $t_f$ may be represented by looking if the paths followed pass through a state correspondig to the projection operator 
\begin{align}
	\mathcal{P}=|f\rangle \langle f|.
	\label{ln155b}
\end{align}
Really such a final state projector can easily be of the type $\mathcal{P}$ projecting on an interval $I$ of dynamical quantities discussed above since typically some variables are measured to be inside small ranges, which we could call $I$. Inserting (\ref{ln155b}) for $\mathcal{P}$ into our fundamental postulate (\ref{ln140}) we obtain 
\begin{align}
	P(|f\rangle )=
	\frac{\displaystyle \sum_{\text{BOUNDARIES}}
	\left|\langle B(t_f)|f\rangle \right|^2
	\left|\langle f|A(t_f)\rangle \right|^2}
	{\displaystyle \sum_{\text{BOUNDARIES}}
	\left|\langle B(t_f)|A(t_f)\rangle \right|^2}.
\end{align}
By insertion of the time development of (\ref{ln155a}) expression 
\begin{align}
	|A(t_f)\rangle =U|A(t_i)\rangle =U|i\rangle
\end{align}
where $U$ is the time development operator from $t_i$ to $t_f$, we get further (ignoring the unimportant sums over boundaries) 
\begin{align}
	P(|f\rangle )=
	\frac{\displaystyle \left|\langle B(t_f)|f\rangle \right|^2
	\left|\langle f|U|i\rangle \right|^2}
	{\displaystyle \left|\langle B(t_f)|U|i\rangle \right|^2}.
	\label{ln157b}
\end{align}
If we allow ourselves to insert here the statistical approximation (\ref{ln142}) we reduce this to 
\begin{align}
	P(|f\rangle )=
	\frac{\displaystyle \left|\langle f|U|i\rangle \right|^2}
	{\displaystyle ||U|i\rangle ||^2}
	\label{ln157c}
\end{align}
using the normalization of $|f\rangle$ already assumed (otherwise $\mathcal{P}=|f\rangle \langle f|$ would not have been a (properly normalized) projection operator). Since we here already assumed $|A(t_i)\rangle =|i\rangle$ i.e. (\ref{ln155a}) this equation (\ref{ln157c}) is precisely the earlier (\ref{ln92}). So our postulate (\ref{ln140}) leads under use of the approximation (\ref{ln142}) to the quite sensible equation (\ref{ln92}).

\subsection{Talk about Feynman diagrams}

Since our model contains the usual theory as the special case of zero imaginary part it needs of course all the usual calculational tricks of the usual theory such as Feynman diagrams to evaluate the S-matrix $U$ giving the time development from $t_i$ to $t_f$.

Since we argued above that the transition from the Feynman path integral to the Hamiltonian formalism for the purposes of obtaining $U$ or the time development of $|A(t)\rangle$ just can be performed by working with the complex coefficients in the Lagrangian, it is not difficult to see that we can also develop the Feynman diagrams just by inserting the complex couplings etc.

One should, however, not forget that our formulae for tha transition probabilities (\ref{ln92}) contains a nontrivial normalization denominator put in to guarantee that the probability assigned to a complete set of states at time $t_f$ summed up be precisely unity. This normalization denominator would be trivial in the usual case of a unitary $U$, but in our model it is important to include it, since otherwise we would not have total probability $1$ everything that could happen at $t_f$ together.

In the usual theorem we have the optical theorem ensuring that the imaginary part of the forward scattering amplitude $\text{Im} T$ is so adjusted as to by interfering with the unscattered beam to remove from the continuing unscattered beam just the number of scattering particles as given by the total cross section $\sigma_{tot}$. When we with our model get a nonunitary $U$ it will typically mean that this optical theorem relation will not be fulfilled. For instance we might fill into the Feynman diagram e.g. a Higgs propagator with a complex mass square 
\begin{align}
	m_H^2=m_{HR}^2+im_{HI}^2
\end{align}
so as to get 
\begin{align}
	prop=\frac{i}{p^2-m_{HR}^2-im_{HI}^2}
\end{align}
for the propagator. That will typically lead to violation of the optical theorem.

Now the ideal momentum eigenstates usually discussed with S-matrices is an idealization and it would be a bit more realistic to consider a beam of particles comming with a wave packet state of a finite area A measured perpendicularly to the beam direction. Suppose that the at first by summing up different possible scatterings gives a formal cross section $\sigma_{tot\> U}$ while the imaginary part of the forward elastic scattering amplitude would correspond via optical theorem to $\sigma_{opt\> U}$. Then the probability for no scattering would if we did not normalize with the denominator be 
\begin{align}
	P_{no\> sc\> U}=\frac{A-\sigma_{opt\> U}}{A}
\end{align}
while the formal total scattering probability would be 
\begin{align}
	P_{sc\> U}=\frac{\sigma_{tot\> U}}{A}.
\end{align}
These two prenormalization probabilities would not add to unity but to 
\begin{align}
	P_{sec\> U}+P_{no\> sec\> U}
	=\frac{\sigma_{tot\> U}-\sigma_{opt\> U}}{A}+1.
\end{align}
That is to say our prediction for scattering would be 
\begin{align}
	P_{sc}
	& =\frac{P_{sc\> U}}{1+\frac{\sigma_{tot\> U}-\sigma_{opt\> U}}{A}} 
	\nonumber \\
	& =\frac{\sigma_{tot\> U}}{A+\sigma_{tot\> U}-\sigma_{opt\> U}} 
	\nonumber \\
	& \simeq \frac{\sigma_{tot\> U}}{A} \qquad 
	(\text{for }A\gg \sigma_{tot\> U},\sigma_{opt\> U})
\end{align}
while the probability for no scattering would be 
\begin{align}
	P_{no\> sc}
	& =\frac{P_{no\> sc\> U}}
	{1+\frac{\sigma_{tot\> U}-\sigma_{opt\> U}}{A}} 
	\nonumber \\
	& =\frac{1-\sigma_{opt\> U}}
	{1+\frac{\sigma_{tot\> U}-\sigma_{opt\> U}}{A}} 
	\nonumber \\
	& =\frac{A-\sigma_{opt\> U}}{A+\sigma_{tot\> U}-\sigma_{opt\> U}} 
	\nonumber \\
	& \simeq 1-\frac{\sigma_{tot\> U}}{A} \qquad 
	(\text{for }A\gg \sigma_{tot\> U},\sigma_{opt\> U}).
\end{align}
From this we see that we would -as we have put into the moodel- see consistency of total number of scatterings and particles removed from the beam; but if we begin to investigate Coulomb scattering interfering with the imaginary part of the elastic scattering amplitude proportional to $\sigma_{opt\> U}$ then deviations from the usual theory may pop up!

\section{Some discussion of the interpretation of our model}

It is obviously of great importance for the viability of our model that the features of a solution decission for whether it is being realized dominantly lie in the era of ``Big Bang" or at least in the past compared to our time. Otherwise we do not have even approximately the rules of physics as we know them, especially the second law of thermodynamics and the fact that we easily find big/macroscopic amounts of a spesial material (but we cannot get mixtures seperate time progresses without making use og other chemicals or free energy sources).

A good hypothesis to arrange such a phenomenologically wished for result would be that the different (possible) solutions -due to the physics of the early, the Big Bang, era have a huge spread in the contribution $S_{I\> BB\> era}=\int_{BB\> era}L_Idt$ from this era. If the variation $\Delta S_{I\> BB\> era}$ of $S_{I\> BB\> era}$ due to varying the solution in the Big Bang era is very big compared to say the fluctuations $\Delta S_{I\> our\> era}$ the contribution $S_{I\> our\> era}$ from our own era then the solution chosen to be realized will be dominantly influenced from what happened at the Big Bang times rather than today. However, it may still be important whether: 
\begin{enumerate}
\item The realized solution is (essentially) \underline{the} one with the smallest possible $S_I$ at all 

or
\item There is such a huge number of solutions to $\delta S_R=0$ and such a huge increase in their number by allowing for somewhat bigger $S_I$ than the minimal $S_I$ one gets so many times more solutions that it overcompensates for the probability (density) factor $e^{-2S_I}$.
\end{enumerate}
In the case 1) of the just the minimal $S_I$ solution being realized the importance of the $S_I$-contributions from the non dominant eras can be almost completely competed out. In the case 2) in which still at first the realized solution is randomly chosen among a very large number of solution it seems unavoidable that among two possible trajectories deviating by a contribution to its $S_I$ by some amount of order unity from todays era -say even a by us understandable contribution of order unity- the one with the smaller (i.e. mere negative) $S_I$ will be appreciably more likely than the other one. This case 2) situation seems to lead to effects that would be very difficult to have got so hidden that nobody had stopped them untill today. At least when working with relativistic particles we would expect big effects in possibility 2) Scattering angle distributions in relativistic scattering processes could be significantly influenced by how long the scattered particle would be 
 allowed to keeps its relativistic velocity after the scattering. If the scattered particle were allowed to escape to outer space we would expect a strongly deformed scattering angle spectrum whereas a rapid stopping of the scattered particles would diminish the deformation of the angular distribution relative to that of the usual real action theory. So it is really much more attractive with the possibility 1) that it is \underline{the} minimal $S_I$ solution which is realized. In that case 2) it would also be quite natural that the contribution from the future to $S_I$ say $S_{I\> future}$ could be quite dominant compared to those being so near in time to today that we have the sufficient knowledge about them to be able to observe any effects.

If there is only of order one or simply just one minimal $S_I$ solution it could be understandable that it were practically fixed by the very strong contributions in the Big Bang era and from random or complicated to evaluate contributions from a very far reaching future era, so that the near to today contributions to $S_I$ would be quite unimportant.

It should be obvious that in the case 1) the effects of practically accessible $S_I$-contributions depending on quantities measured in an actual experiment will be dominated out so that they will be strongly suppressed, they will drown in for us to see random contributions from the future or the more organized contributions from Big Bang time determining the initial state. In the classical approximation the Big Bang initial time contributions to $S_I$ may be all dominant, but quantum mechanically we have typically a prepared state $|i\rangle$ which will be given by the Big Bang era $S_I$-contributions while the measurement of a final state in principle could be more sensitive to the future $S_I$-contributions.

Now, however, under the possibility 1) of the single solution being picked with the totally minimal $S_I$ the state $|B(t)\rangle$ will most likely be dominantly determined from the far future and will have compared to that very little dependence on the $S_I$-contributions of the near future (a rather short time relative to the far future). By the argument that the real part $S_R$ courses there to be a complicated development through time -which we take to be ergodic- one thus obtains $|B(t)\rangle$ up to an overall scale becomes a random state selected with same probability among all states in the Hilbert space. In this case 1) situation we thus derive rather convincingly in the ergodicity-approximation that we can indeed approximate 
\begin{align}
	\frac{|B(t)\rangle \langle B(t)|}{\langle B(t)|B(t)\rangle}
	\sim {\bf 1}.
\end{align}
Really it is better to think forward to a moment of time say $t_{erg}$ which is on the one hand early enough that we can use the ``ergodicity-approximation" and on the other hand late enough that the $L_I$'s later are in pracsis zero over the time scales of the experiment so that $H$ from $t_{erg}$ on can be counted practically Hermitean. This should mean that at that time $t_{erg}$ the system has fallen back to usual states in which $L_I$ is trivial. Then at $t=t_{erg}$ we simply have (\ref{ln142}) and we have it for $t$ later than $t_{erg}$ i.e. in pracsis 
\begin{align}
	|B(t)\rangle \langle B(t)|\propto {\bf 1}.
\end{align}
for $t\geq t_{erg}$. Then even this approximation (\ref{ln142}) is selfconsistent for the later than $t_{erg}$ times. This selfconsistency of having (\ref{ln142}) at one moment of time is not true over time time intervals over which we do not effectively have a Hermitean Hamiltonian $H$.

\subsection{Development to final S-matrix}

We may now develop formula (\ref{ln157b}) by using instead of $|B(t_f)\rangle$ a $B$-state for a time later than $t_{erg}$ or we simply use $|B(t_{erg})\rangle$ just at the time $t_{erg}$. Then the transition probability from $|i\rangle$ to $|f\rangle$, i.e. (\ref{ln157b}) becomes 
\begin{align}
	P(|f\rangle ) & =
	\frac{\left|\langle B(t_{erg})|U_{t_f\to t_{erg}}|f\rangle\right|^2
	\cdot \left|\langle f|U|i\rangle\right|^2}
	{\left|\langle B(t_{erg})|U_{t_f\to t_{erg}}U|i\rangle\right|^2} 
	\nonumber \\
	& =\frac{\left|\left|U_{t_f\to t_{erg}}|f\rangle\right|\right|^2
	\cdot \left|\langle f|U|i\rangle\right|^2}
	{\left|\left|U_{t_f\to t_{erg}}|i\rangle\right|\right|^2}
	\label{nb192}
\end{align}
where we have defined the time development operator $U_{t_f\to t_{erg}}$ performing with the nonhermitean Hamiltonian $H$ the development from $t_f$ to the even later time $t_{erg}$. We also defined the analogous development operator from the initial time $t_i$ of the experiment to the time $t_{erg}$ from which we practically can ignore $L_I$ to be called $U_{t_i\to t_{erg}}$. So we have since really analogously $U=U_{t_i\to t_f}$, that 
\begin{align}
	U_{t_i\to t_{erg}}=U_{t_f\to t_{erg}}U.
\end{align}
We could now simply (\ref{nb192}) by introducing the states $|f\rangle$ and $|i\rangle$ refered by time propagation to the time $t_{erg}$ by defining 
\begin{align}
	|f\rangle_{erg} & =U_{t_f\to t_{erg}}|f\rangle ; \nonumber \\
	|i\rangle_{erg} & =U_{t_i\to t_{erg}}|i\rangle \nonumber \\
	& =U_{t_f\to t_{erg}}U|i\rangle .
\end{align}
In these terms we obtain the $|i\rangle$ to $|f\rangle$ transition probability (\ref{nb192}) developped to 
\begin{align}
	P(|f\rangle ) 
	&=\frac{\big|\big||f\rangle_{erg}\big|\big|^2\left|\langle f|_{erg}
	(U_{t_f\to t_{erg}}^{-1})^{\dagger}U_{t_f\to t_{erg}}^{-1}
	|i\rangle_{erg}\right|^2}{\big|\big||i\rangle_{erg}\big|\big|^2} 
	\nonumber \\
	&=\frac{\big|\big||f\rangle_{erg}\big|\big|^2\big|\langle f|_f
	|i\rangle_f\big|^2}{\big|\big||i\rangle_{erg}\big|\big|^2} 
	\nonumber \\
	&=\frac{f\big|\langle f|_f|i\rangle_f\big|^2}{i}, 
	\label{nb196}
\end{align}
we have defined 
\begin{align}
	f=\frac{\big|\big||f\rangle_{eng}\big|\big|^2}
	{\big|\big||f\rangle_f\big|\big|^2}
	=\frac{\big|\big||f\rangle_{eng}\big|\big|^2}
	{\big|\big|U_{t_f\to t_{eng}}^{-1}|f\rangle_{eng}\big|\big|^2}
	\label{nb197a}
\end{align}
and 
\begin{align}
	i=\frac{\big|\big||i\rangle_{eng}\big|\big|^2}
	{\big|\big||i\rangle_f\big|\big|^2}
	=\frac{\big|\big||i\rangle_{eng}\big|\big|^2}
	{\big|\big|U_{t_f\to t_{eng}}^{-1}|i\rangle_{eng}\big|\big|^2}
	\label{nb197b}
\end{align}
Obviously we would for normalized $|i_j\rangle_f$ and $|f_k\rangle_f$ basis vectors of respectively a set involving $|i\rangle_f$ and $|f\rangle_f$ have that the matrix $\langle f_j|_f|i_k\rangle_f$ is unitary.

This means that our expression (\ref{nb196}) so far got of the form of a unitary -almost normal- S-matrix modified by means of the extra factor $i$ and $f$ depending only on the initial and final states respectively (\ref{nb197a}),(\ref{nb197b}).

\section{Expected effects of performing the measurement}

Normally in quantum mechanics the very measurement process seems to play a significant role.

Here we shall argue that depending upon whether our model is working in the case 1) of the solution path with the absolutely minimal $S_I$ or in the case 2) of there being so many more solutions with a somewhat higher $S_I$ that statistically one of these not truly minimal $S_I$ solutions become the realized one the performance of the measurement process come to play a role in our model. Indeed we shall argue that the first derived formulas such as (\ref{nb196}) e.g. are only true including the measurement process in the case 2) of a not truly minimal $S_I$ solution being realized. In the case 1) we shall however see that almost all effects of our imaginary part model disappears.

\subsection{The extra and random contribution to $S_I$ depending on the measurement result}

We want to argue that we may take it that we obtain an extra contribution to $S_I$ effectively depending on the measurement result. This contribution we can take to be random but with expectation value zero, so that we think of a pure fluctuation.

In order to argue for such a fluctuating contribution let us remark that a very important feature of a quantum mechanical measurement is that it is associated with an enhancement mechanism. That could e.g. be the crystalization of some material caused by a tiny bit of light -a photon- or some other single particle. A typical example could be the bubble in the bubble chamber also. Again a single particle passing through or a single electron excited by it causes a bubble containing a huge number of particles to form out of the overheated fluid. We can call such processes enhancements because they have a little effect cause a much bigger effect and that even a big effect of a regular type. We can give a good description of the bubble in simple words, it is not only as the butterfly in the ``butterfly effect" which also in the long run can cause big effects. The latter ones are practically almost uncalculable, while the bubble formation caused by the particle in the bubble chambe
 r is so well understood that we use it to effectively ``see" the particles.

If the measurement is made to measure a quantity $O$, say, it measures that these regular or systematic enhancement effects reflect the value of $O$ found. Otherwise it would not be a measurement of $O$. The information of the then not measured conjugate variable of $O$ is not in the same way regularly or systematically enhanced. So it is the $O$-value rather than the value of its canonical conjugate variable $\Pi_O$ which gets enhanced.

Thus the consequencies for the future of the measurement time $t_{measurement}$ and thus for the $S_I$-contribution 
\begin{align}
	S_{I\> fut.\> mes.}=\int_{t_{measurement}}^{\infty}L_Idt
	\label{nb208}
\end{align}
also depend on $O$ (rather than on $\Pi_O$). It is this contribution -which is of course at the end in general very hard to compute- that we want to consider as practically random numbers depending on the $O$-value measurement (in the measurement considered). We must imagine that the various consequencies of the measurement result of measurement of $O$ becomes at least a macroscopic signal in the electronics or the brain of the experimentor, but even very likely somehow come to influence a publication about that experiment. So in turn it influences history of science, history of humans, and at the end all of nature. That cannot avoid meaning that we have a rather big $O$-dependent contribution $S_{If}(O^{\prime})$ to the integral (\ref{nb208}), where $O^{\prime}$ is the measure value of $O$.

\subsection{Significance of case 1) versus case 2)}

We may a priori expect that such $O$-dependent contributions $S_{If}(O^{\prime})$ being integrals of the whole future of essentially macroscopic contributions will be much bigger than the contribution to $S_I$ comming from the very quantum experiment during the relatively much smaller time span $t_f-t_i$ during which, say, the scatterings or the like takes place. The only exception would seem to be if there were some parameter -the imaginary part $m_{HI}^2$ of the Higgs mass square, say- which were much larger in order of magnitude than the usual contributions to $S_I$ and especially to the $S_{If}(O^{\prime})$'s. Baring such an enormous contribution during the short period $t_i$ to $t_f$ it is then $S_{If}(O^{\prime})$ that dominates over the $S_I$-contributions from the ``scattering time".

\subsection{Case 2): A solution among many}

If our model is working in the case 2) of it being not the very minimal $S_I$ solution that get realized but rather a solution belonging to a much more copiously populated range in $S_I$ that gets realized we should really not call our $O$-value dependent contribution just $S_{If}(O^{\prime})$, but rather $S_{If}(O^{\prime},sol)$. By doing that we should namely emphasize that we obtain (in general) completely different contributions depending on the $O$-value depending on which one solution among the many solutions gets realized.

In this case 2) it is in fact rather the average over the many possible solutions of $e^{-2S_I}$ evaluated under the restriction of the various $O$-values being realized that gives the probabilities for these various $O$-values $O^{\prime}$. Since we would, however, at least assume as our ansatz that the $S_{If}(O^{\prime})$ distributions are the same for the various $O$-values the relative probabilities of the various measurement results $O^{\prime}$ or $O^{\prime \prime}$ for the quantity $O$ will not be much influenced by $S_{If}(O^{\prime},sol)$ contributions. Thus the relatively small contribution from the short $t_i$ to $t_f$ time of just a few particles may have a chance to make themselves felt and formulas derived with such effects are expected to be o.k. in this case 2).

\subsection{Case 2): Absolute minimal $S_I$ solution realized}

In the case 1) on the other hand of only one absolutely minimal $S_I$ solution being realized we do not have to worry so much about the dependence on the many different solutions with a given $O$-value $O^{\prime}$ because there is only the one with the minimal $S_I$ which has a chance at all. In this case 1) we could imagine that $S_{If}(O^{\prime})$ should be defined as corresponding the (classical) solution going through $O=O^{\prime}$ and having among that class of solutions the minimal $S_I$. Since the contribution $S_{If}(O^{\prime})$ is still much bigger than the $S_I$-contribution from scattering experiment say (i.e. from the $t_i$ to $t_f$ experiment considered) we see that in this case 1) the contributions from the experiment time proper has no chance to come through. The effects will almost completely drown in the effectively random $S_{If}(O^{\prime})$ contribution.

That should make it exceedingly difficult to see any effects at all of our imaginary action $S_I$ model in this case 1) That makes our case 1) very attractive phenomenologically, because after all no effects of influence from the future has been observed at all so far.

\section{How to correct our formulas for case 1)?}

The simplest way to argue for the formalism for the case 1) of a single minimal -\underline{the} lowed $S_I$ one- solution being realized is to use the classical approximation in spite of the fact that we are truly wanting to consider quantum experiments: If you imagine almost the whole way a classical solution which has the measured value $O$ being $O^{\prime}$ and in addition having the minimal $S_I$ among all the solutions with this property solution which has the measured value $O$ being $O^{\prime}$ and in addition having the minimal $S_I$ among all the solutions with this property and call the $S_I$ for that solution $S_{I\> min}(O^{\prime})$ then the realized solution under assumption of our case 1) should be that for the $O^{\prime}$ which gives the minimal $S_{I\> min}(O^{\prime})$.

Now the basic argumentation for the $S_I$-contribution $S_{I\> during\> exp}$ comming from the particles in the period in which they are considered scattering particles and described by an S-matrix $U$ being unimportant goes like this: Imagine that we consider two different calculations, one a) in which we just formally switched these ``during S-matrix" contributions $S_{I\> during\> exp}$ off, and one b) in which these contributions $S_{I\> during\> exp}$ are included. If the differences between the various $S_{I\> min}(O^{\prime})$ (for the different eigenvalues of the measured quantity $O$, called $O^{\prime}$) deviate by amounts of imaginary action much bigger than the contribution $S_{I\> during\> exp}$ i.e. if 
\begin{align}
	S_{I\> during\> exp}\ll S_{I\> min}(O^{\prime})
	-S_{I\> min}(O^{\prime \prime})
	\label{nb226}
\end{align}
typically, then the switching on of the $S_{I\> during\> exp}$-contribution, i.e. going from a) to b) will only with very little probability cause any change in which of the $S_{I\> min}(O^{\prime})$ imaginary action values will be the minimal one. Thus under the assumption (\ref{nb226}) the switching on or off of $S_{I\> during\> exp}$ makes only little difference for the measurement result $O^{\prime}$ and we can as well use the calculation a) with the $S_{I\> during\> exp}$ switched off. But that means that provided (\ref{nb226}) we can in the S-matrix formulas totally leave out the imaginary action.

This is of course a very important result which as we stressed already only apply in the case 1) that the single minimal $S_I$ classical solution with the absolutely minimal $S_I$ is the one chosen. Really it means that there must not be so many with higher (but perhaps numerically smaller negative) $S_I$-values that the better chance of one of these higher $S_I$ ones can compensate for the weight factor $e^{-2S_I}$.

\subsection{A further caveat}

There is a slightlydifferent way in which the hypothesis -case 1)- if only the single classical solution with minimal $S_I$ being realized may be violated:

There might occur a quantum experiment with a significant interference between two different possible classical solutions. One might for instance think of the famous double slit experiment discussed so much between Bohr and Einstein in which a particle seems to have passed through two slits, without anybody being able to know which without converting the experiment into a different one. In order to reproduce the correct interferences it is crucial that there are more than one involved classical path. This means that for such interference experiments the hypothesis of case 1) of only one classical path being realized is formally logically violated. It is, however, in a slightly different way than what we described as the case 2) Even if we have to first approximation case 1) in the sense that all over except for a few short intervals in time there were only one single classical solution selected, namely the one with minimal $S_I$, an even not so terrible big $S_{I\> during\> e
 xp}$ contribution being different for say the passage through the two different slits in the double slit experiment could cause suppression of the quantum amplitude for one of the slits relative to that for the other one. Such a suppression would of course disturb the interference pattern and thus cause an in general observable effect. The reason that our argument for no observable effects of $S_{I\> during\> exp}$ in case 1) does not work in the case here of two interfering classical solutions is that both of them ends up with the same measured value $O^{\prime}$ of $O$. Then of course it does not make any difference if the typical difference $S_{I\> min}(O^{\prime})-S_{I\> min}(O^{\prime \prime})$ is big or not. Well, it may be more correct to say that the difference in $S_I$ for two interfering paths in say a double slit is only of the order of $S_{I\> during\> exp}$ and we must include in our calculation both paths if there shall be interference at all. Ever so big $S_I$
 -contributions later in future cannot distinguish and choose the one path relative to another one interfering with it. If there were effects of which of the interfering paths had been chosen in the future -so that they could give future $S_I$-contributions- it would be like the measurements of which path that are precisely impossible without spoiling the experiments with its interference.

\subsection{Conclusion about our general suppression of $S_I$-effects}

The above discussion means that under the hypothesis that apart from paths interfering the realized classical solution or more precisely the bunch of effectively realized interfering classical solutions is uniquely the bunch with minimal $S_I$, essentially case 1), we cannot observe $S_{I\> during\> exp}$, \underline{except} through the modification it causes in the interference.

Apart from this disturbance of interference patterns the effect of $S_I$ concerns the very selection of the classical path realized, but because of our accessible time to practically observe corrections is so short compared to future and past such effects are practically negligible, except for how they might have governed cosmology.

On top of these suppression of our $S_I$-effects to only occur for interference or for cosmologically important decissions we have that massless or conserved nonrelativistic particles course (further) suppression.

So you would actually have the best hopes for seeing $S_I$-effects in interference between massive particles on paths deviating with relativistic speeds or using nonconserved particles such as the Higgs particle.

Actually we shall argue that using the Higgs particles is likely to be especially promissing for observing effects of the imaginary part of the action. It is not only that we here have a particle with mass different from zero which is not conserved, but also that it well could be that the imaginary part of the Higgs mass square $m_I^2$ were exceptionally big compared to other contribution at accessible energy scales. The point is that it is related to the well known hierarchy problem that the Higgs real mass square term is surprisingly small. Privided no similar ``theoretical surprise" makes also the imaginary part of the Higgs mass square small, size of $m_I^2$ would from the experimental scales point of view be tremendously big.

One of the most promissing places to look for imaginary action effects is indeed suggested to be where one could have interence between paths with Higgs particles existing over different time intervals. This sort of interference is exactly what is observed if one measures a Higgs particle Breit-Wigner mass distribution by measuring the mass of actual Higgs particle decay products sufficiently accurately to evaluate the shape of the peak. In the rest of the present article we shall indeed study what our imaginary action model is likely to suggest as modification of the in usual models expected Breit-Wigner Higgs mass distribution. We shall indeed suggest that in our model there will likely be a significantly broader Higgs mass distribution than expected in conventional models. In fact we at the end argue for a Higgs mass distribution essentially of the shape of the \underline{square root} of the usual Breit-Wigner distribution. It means it will fall off like $\frac{1}{|M-M_{Hi
 ggs}|}$ for large $M-M_{Higgs}$ rather than as $\frac{1}{|M-M_{Higgs}|^2}$ as in the usual theory.

\section{Higgs broadening}

It is intuitively suggestive that if Higgs particles so disfavours a solution to be realized that Higgs production is supressed if needed by almost miraculous events then we would also not expect the Higgs to live the from usual physics expected lifetime. If so then the Higgs would get its average lifetime reduced and by Heisenberg uncertainty relation one would expect also a broader width for the Higgs than usual.

But is this true, and how can we estimate how broad?

At first one would presumably have expected that the timedependence form 
\begin{align}
	\propto e^{-\frac{\Gamma}{2}t} \qquad 
	\text{for }t\geq t_{\text{HIGGS CREATION}}
\end{align}
(let us put $t_{\text{HIGGS CREATION}}$=0) would be modified simply by being multiplied by the square root of the probability supression factor $e^{-2\frac{m_{HI}^2}{2m_{HR}}t}$ so that we end up with the total decay form of the amplitude -for the Higgs still being there- 
\begin{align}
	e^{\left(-\frac{\Gamma}{2}-\frac{m_{HI}^2}{2m_{HR}}\right)t}
\end{align}

This formula seemingly representing an effective decay rate of the Higgs can, however, hardly be true, because the Higgs can only decay -even effectively- once it is produced provided it has something to decay to and decays sufficiently strongly. The correction for this must come in via a normalization taking care of that once we got the Higgs produced in spite of its ``destination" of bring long and know that then we have to imagine that since it happened the $S_I$ contribution from the past were presumably relatively small so as to compensate for the effect of the Higgs long life. If really we are allowed to think about a specific decay moment for the Higgs, then we should presume that the extra contribution $\frac{m_{HI}^2}{2m_{HR}}\tau$ to $S_I$ from a Higgs living the eigentime $\tau$ would if it were known to live so would have been cancelled by contributions from before or after. Thus at the end it seems that the whole effect is cancelled if we somehow get knowing how 
 long the Higgs lives. Now, however, the typical situation is that even if by some coproduction we may know that a Higgs were born, then it will decay usually so that during the decay process it will be in a superposition of having decayed and having not yet decayed. One would then think that provided we keep the state normalized -as we actually have to since in our model there is probability unity for having a future- it is only when there is a significant probability for both that the Higgs is still there and that it is already decayed in the wave function there is a possibility for the imaginary Lagrangian $L_I$ contribution to make itself felt by increasing the probability for the decay having taken place.

A very crude estimate would say that if we denote the width $\Gamma_{SI}=\frac{m_{HI}^2}{m_{HR}}$ and take it that $\Gamma_{SI}\gg \Gamma_{b\bar{b}}$ (the main decay say it were $b\bar{b}$) corresponding to the probability decay then we need that the probability for the decay into say a main mode $b\bar{b}$ has already taken place to be of the order $\frac{\Gamma}{\Gamma_{SI}}$ if we shall have of order unity final disappearance of the Higgs particle.

That might be the true formula to Fourier transform to obtain the Higgs width broadening if it were not for the influence from the future also built into our model. In spite of the fact that a short life for a certain Higgs of course would -provided as we assume $m_{HI}^2>0$- contribute to make bigger the likelyhood of a solution with such a short Higgs life, it also influences what goes on in the future relative to the Higgs decay. In this comming time the exact value of the Higgs lifetime in question will typically have very complicated and untransparent but also very likely big effects on what will go on. Thus if there are just some $S_I$-contribution in this future relative to the Higgs decay the total $S_I$-contribution may no longer at all be a nice linear function of the eigenlife time $-\frac{\tau m_{HI}^2}{2m_{HR}}$ but likely a very complicated strongly oscillating function. Now these contributions comming -as we could say- from the ``butterfly wing effect" of the H
 iggs lifetime for the Higgs in question could easily be very much bigger than the contribution from the Higgs living only of the order of $\frac{1}{\text{MeV}}\sim 10^{-21}\text{s}$ say, since the the presumed yet to exist time of the Universe could e.g. be of the order of tens of millions (i.e. $\sim 10^{10}$) years. Even if for the reason of the imaginary part of the Higgs mass square not being by the solution to hierarchy problem mechanism suppressed like the real one would be say $10^{34}$ times bigger than the real part, it is not immediately safe whether it could compete with the contribution from the long times. At least unless the imaginary part $m_{HI}^2$ of the coefficient in the Higgs mass term in the Lagrangian density is compared to the real part abnormously large the contributions to $S_I$ from the much longer future than the Higgs lifetime order of magnitude will contribute quite dominantly compared to the contribution from the short Higgs existence to $S_I$. 
 Under the dominance of the future contribution the $S_I$ favoured Higgs lifetime could easily be shifted around in a way we would consider random. In fact the future $S_I$-contribution will typically depend on some combination of the Higgs lifetime $\tau$ and its conjugate variable its rest energy $m$. Thereby the dominant $S_I$-contribution could easily be obtained for there being a lifetime wave function $\psi_{life}(t)$ which has a distribution in the lifetime $t_{life}$.

We may estimate the effective Higgs width after the broadening effect in a couple of different ways: 

\subsection{Thinking of a time development}

In the first estimate we think of a Higgs being produced at some time in the rest frame $\tau =0$ say. Now as time goes on -in the beginning- there grows in the usual picture, and also in ours in fact an amplitude for this initial Higgs having indeed decayed into say $|b\bar{b}\rangle$ a state describing a $b$ and an anti $b$-quark $\bar{b}$ having been produced. The latter should in the beginning come with a probability $\langle b\bar{b}|b\bar{b}\rangle \propto \tau \Gamma_{USUAL}$ where $\tau$ is the here assumed small eigentime passed since the originating Higgs were produced and $\Gamma_{USUAL}$ is the in the usual way calculated width of the Higgs particle (imagined here just for pedagogical simplicity to be to the $b\bar{b}$-channel).

In the full amplitude/state vector for the Higgs or decay product system after time $\tau$ we have the two terms 
\begin{align}
	|full\rangle &=|\mathcal{H}\rangle +|b\bar{b}\rangle \nonumber \\
	&=\alpha |\mathcal{H}\rangle_{norm}+\beta |b\bar{b}\rangle_{norm}
	\label{bphwe4}
\end{align}
where $|\mathcal{H}\rangle_{norm}$ is a to unit norm, $\langle \mathcal{H}|_{norm}|\mathcal{H}\rangle_{norm}=1$ normalized Higgs particle while $|b\bar{b}\rangle_{norm}$ is the also normalized appropriate $b\bar{b}$ state. The symbols 
\begin{align}
	|\mathcal{H}\rangle =\alpha |\mathcal{H}\rangle_{norm}
\end{align}
and 
\begin{align}
	|b\bar{b}\rangle =\beta |b\bar{b}\rangle_{norm}
\end{align}
on the other hand stands for the two parts of the full amplitude or state (\ref{bphwe4}). 

Now if we take it that -by far- the most important part of the imaginary part of the Lagrangian i.e. $L_I$ is the Higgs mass square term (remember the argument that a solution to the large weak to Planck scale ratio could easily solve this problem for the real part of the mass square $m_{HR}^2$ while leaving the imaginary part untuned and thus large from say L.H.C. physics scale) then as $\tau$ goes on the probability for the $|\mathcal{H}\rangle$ part of the amplitude surviving should for the $S_I$-effect reason fall down exponentially. That is to say that we at first would say that the amplitude square for this survival $|\alpha (\tau )|^2$ should fall off exponentially like 
\begin{align}
	|\alpha (\tau )|^2 &\propto \exp \left(-2S_I\big|_{Higgs}\right) 
	\nonumber \\
	&\simeq \exp \left(-2L_I\big|_{Higgs}\tau \right) \nonumber \\
	&\simeq \exp \left(-\frac{|m_{HI}^2|}{m_{HR}}\tau \right)
\end{align}
(for the assumed ``habed" Higgs spin). This can, however, not be quite so simple since we must the normalization conserved to unity meaning 
\begin{align}
	|\alpha (t)|^2+|\beta (t)|^2=1.
	\label{bphwe10}
\end{align}
This equation has to be uphold by an overall normalization. In the situation in the biginning, $\tau$ small, and assuming that the ``Imaginary action width" 
\begin{align}
	\Gamma_{SI} &\hat{=}2\left|L_I\big|_{Higgs}\right| \nonumber \\
	&=\frac{|m_{HI}^2|}{|m_{HR}|}\gg \Gamma_{USUAL}
\end{align}
we at first would expect 
\begin{align}
	& |\beta (t)|^2 \propto \Gamma_{USUAL}\tau \nonumber \\
	& |\alpha (t)|^2\propto \exp (-\Gamma_{SI}\tau ),
\end{align}
but then we must rescale the normalization to ensure (\ref{bphwe10}). Then rather 
\begin{align}
	& |\alpha (t)|^2\simeq \frac{\exp (-\Gamma_{SI}\tau )}
	{\exp (-\Gamma_{SI}\tau )+\Gamma_{USUAL}\tau} 
	\nonumber \\
	& |\beta (t)|^2 \simeq \frac{\Gamma_{USUAL}\tau}
	{\exp (-\Gamma_{SI}\tau )+\Gamma_{USUAL}\tau}.
\end{align}
Inspection of these equations immediately reveals that there is no essential decay away of the (genuine) Higgs particle 
\begin{align}
	\langle \mathcal{H}|\mathcal{H}\rangle =|\alpha (\tau )|^2
\end{align}
before the two terms in the normalization denominator 
\begin{align}
	\exp (-\Gamma_{SI}\tau )+\Gamma_{USUAL}\tau 
\end{align}
reach to become of the same order. In the very beginnig of course the term $\exp (-\Gamma_{SI}\tau )\sim 1$ dominates over the for small $\tau$ small $\Gamma_{USUAL}\tau$. But once this situation of the two terms being comparable the Higgs particle essentially begins to decay. If we therefore want to estimate a very crude effective Higgs decay time in our model 
\begin{align}
	\tau_{eff}=\frac{1}{\Gamma_{eff}}
\end{align}
this effective lifetime $\tau_{eff}$ must be crudely given by 
\begin{align}
	\Gamma_{USUAL}\tau_{eff}\simeq \exp (\Gamma_{SI}\tau_{eff}).
	\label{bphwe16}
\end{align}
From this equation (\ref{bphwe16}) we then deduce after first defining 
\begin{align}
	X\hat{=}\Gamma_{SI}\tau_{eff} \quad \text{or} \quad 
	\tau_{eff}\hat{=}\frac{X}{\Gamma_{SI}}
	\label{bphwe16second}
\end{align}
that 
\begin{align}
	e^{-X}=\frac{\Gamma_{USUAL}}{\Gamma_{SI}}X
\end{align}
and thus ignoring the essential double logarithm $\log X$ that 
\begin{align}
	X\simeq \log \frac{\Gamma_{SI}}{\Gamma_{USUAL}}.
	\label{bphwe17}
\end{align}
Inserting this equation (\ref{bphwe17}) into the definition (\ref{bphwe16second}) of $X$ we finally obtain 
\begin{align}
	\tau_{eff}=\frac{1}{\Gamma_{SI}}X=\frac{1}{\Gamma_{SI}}
	\log \frac{\Gamma_{SI}}{\Gamma_{USUAL}}
\end{align}
or 
\begin{align}
	\Gamma_{eff}
	=\frac{\Gamma_{SI}}{\log \frac{\Gamma_{SI}}{\Gamma_{USUAL}}}.
	\label{bphwe18}
\end{align}
Thus the effective width $\Gamma_{eff}$ of the Higgs which we expect from our model to be effectively seen in the experiments when the Higgs will be or were found (in L.E.P. we actually think it were already found with the mass $115$ GeV) we expect to be given by (\ref{bphwe18}).

Of course we do not really know $m_{HI}^2$ and thus $\Gamma_{SI}=\left|\frac{m_{HI}^2}{m_{HR}}\right|$ to insert into (\ref{bphwe18}), but we may wonder if in the case that $\Gamma_{SI}$ were much longer than the Higgs mass really should replace it by this Higgs mass instead? The reason is that it sounds a bit ccrazy to expect an energy distribution in a resonance peak to extend essentially into negative energy for the produced particle as a width broader than the mass would correspond to.

We could take this as a suggestion to in practice anyway -even if indeed $\Gamma_{SI}$ were even more big- to take $\Gamma_{SI}\sim m_{HR}$. We might also suggest the speculation in the conjugate way of saying: We can hardly in quantum mechanics imagine a start for the existence of a Higgs particle to be so well defined that the energy of this Higgs particle if using Heisenberg becomes so uncertain that it has a big chance of being negative.

Also if $m_{HI}^2$, the imaginary part by some hierarchy problem related mechanism were tuned to the same order of magnitude as $m_{HR}^2$ we would get (in pracsis) $\Gamma_{SI}\sim m_{HR}$. With this suggestion inserted our formula (\ref{bphwe18}) gets rewritten into 
\begin{align}
	\Gamma_{eff}\simeq \frac{m_{HR}}{\log \frac{m_{HR}}{\Gamma_{USUAL}}}
\end{align}
wherein we for orientation may insert the L.E.P. uncertain finding $|m_{HR}\simeq 115|$GeV and a usual decay rate for such a very light Higgs of order of magnitude $\Gamma_{USUAL}\simeq 10^{-3}$GeV. This would give 
\begin{align}
	\Gamma_{eff} &\simeq \frac{115\text{GeV}}{\log 10^5}
	\simeq \frac{115}{23.5}\text{GeV} \nonumber \\
	&\simeq 10\text{GeV}.
\end{align}
This is just a broadening of the Higgs width of the order of magnitude which was extracted from the L.E.P. data to support of the theory of the Higgs mixing with Kaluza-Klein type models. What seems to be in the data is that were statistically more Higgs candidates even below the now efficial lower bound for the mass $144$GeV, and that there has at times been even some findings below with insufficient statistics. The suggestion of the present article of course is that these few events were due to ``broadening" of the Higgs simply indeed Higgs particles. There were even an event with several GeV higher mass than the ``peak" at $115$GeV, but the kinematics at L.E.P. were so that there were hardly possibilities for higher mass candidates.

\subsection{Method with fluctuating start for Higgs particle}

The second method -which we also can use to obtain an estimate of the in our model expected Higgs peak shape- considers it that the Higgs particle is not necessarily created effectively in a fixed time state, but rather in some (linear) combination the energy and the start moment time.

In pracsis the experimentalist neither measures the start moment time nor the energy or mass at the start of the Higgs particle life very accurately compared to the scales needed.

Let us first consider the two extreme cases: 
\begin{enumerate}
\item The Higgs were started (created) with a completely well defined energy (say in some coproduction with the energy of everything else in addition to the beams having measured energies).
\item The moment of creation were measured accurately.
\end{enumerate}
Then in both cases we consider it that we measure the energy or mass rather of the decay products, say $\gamma +\gamma$ or $b+\bar{b}$ accurately.

Now the question is what amount of Higgs-broadening we are expected to see in these two cases: 
\begin{enumerate}
\item Since we simplify to think of only the one degree of freedom -the distance of the say $b+\bar{b}$ from each other- the determination of the energy (or mass) in the initial state means fully fixing the system and the energy at the end is completely guaranteed. So by the initial energy the final is also determined and must occur with probability unity under the preassumption of the initial one. It does not mean, however, as is well known from usual real action theory that there is no Breit-Wigner peak. It is only that if one measures the mass or energy twise, then one shall get the same result both ways: decay product mass versus production mass.

However, concerning our potential modification by the $S_I$-effect it gets totally normalized away in this case 1) of the energy or mass being doubly measured. That should mean that there should be no ``Higgs-broadening" when one measures the mass doubly i.e. both before and after the existence time of the Higgs.

The reason for this cancellation is that with the measurement of the correct energy once more the matrix element squared ratio can be complemented by adding similar terms with the now energy/mass eigenstate $|f\rangle$ replaced by the other non-achievable -by energy contribution- energy eigenstate. Because of the energy eigenstates other than the measured one $|f\rangle$ cannot occur the here proposed to be added terms are of course zero and it is o.k. to add them. After this addition and using that the sum 
\begin{align}
	\sum_{E}|E\rangle \langle E|=\underline{1}
\end{align}
over the complete set of energy eigenstates is of course the unit operator \underline{$1$} we see that numerator and denominator becomes the same (and we are left with only the IFFF factor, which we ignore by putting it to unity). Thus we get in this case of initial energy fixation no $S_I$-effect.
\item In the opposite case of a prepared starting time corresponding to the in pracsis unachievable measurement of precisely \underline{when} the Higgs got created we would find in the usual case the energy i.e. mass (in Higgs c.m.s.) distribution by Fourier transforming the exponential decay amplitude $\propto \exp \left(-\frac{\Gamma_{USUAL}}{2}\tau\right)$. Now, however, one must take into account that this amplitude is further suppressed the larger the Higgs existence time $\tau$ due to the theory caused extra term in the imaginary part of the action $S_I$ 
\begin{align}
	S_I\Big|_{FROM\> HIGGS\> LIFE}=\frac{\Gamma_{SI}}{2}\tau .
\end{align}
This acts as an increased rate of decay of the Higgs particle so as if it had the total edcay rate $\Gamma_{SI}+\Gamma_{USUAL}$. If $\Gamma_{SI}\gg \Gamma_{USUAL}$ that of course means a much broader Higgs Breit-Wigner peak.

Presumably though we should to avoid the problem with negative energies of the Higgs particle only take a total width up to of the order of the Higgs (real) mass $m_{HR}$ seriously. So as under 1) we suggest to in pracsis just put $\Gamma_{SI}\sim m_{HR}\sim 115$GeV say.
\end{enumerate}
But now in practical experiment presumably both the mass prior to decay and the starting time are badly measured compared to the need for our discussions 1) and 2) above. We should, however, now have in mind that it is a characteristic feature of our imaginary action model that the future acts as a kind of hidden variable machinery implying that at the end everything get -by ``hind sight"- fixed w.r.t. mostly both a variable and its conjugate. Really it means that at the end our imaginary action effectively makes a preparation of the Higgs state at production as good as it is possible according to quantum mechanics. This is to be understood that sooner or late and in the past and/or in the future the different recoil particles coproduced with the Higgs as well as the beam particles get either their positions or their momenta or some contribution there of fixed in order to minimize $S_I$. Such a fixation by $S_I$ in past or in future becomes effectively a preparation of the Hi
 ggs state produced. Now it is, however, not under our control -in the case we did not ourselves measure- what was measured the start time or the start energy or some combination?

Most chance there is of course for it being some combination of energy (i.e. actual mass) and the start time which is getting ``measured" effectively by our $S_I$-effects. Especially concerning the part of this ``measurement" that is being determined from the future $S_I$-contribution the ``measurement" finally being done by the $S_I$ at a very late time $t$ it will have been canonically transformed around, corresponding to time developments over huge time intervals. Such enormous transformations canonical transformations in going from what $S_I$ effectively depends on to what becomes the initial preparation setting of the Higgs initial state in question means that the latter will have smeared out by huge canonical transformations. We shall take the effect of the huge or many canonical transformations which we are forced to consider random to imply that the probability distribution for the combinations of starting time and energy that were effectively ``measured" or prepared 
 by the $S_I$-effects should be invariant under canonical transformations. Such a canonical transformation invariant distribution of the combination quantity to be taken as measured seems anyway a very natural assumption to make. Our arguments about the very many successive canonical transformations needed to connect the times at which the important $L_I$-contributions come to the time of the Higgs state being delivered were just to support this in any case very natural hypothesis of a canonically invariant distribution of the combination which say linearized would be 
\begin{align}
	aH_{Higgs}+bt_{start}.
\end{align}
Here $a$ and $b$ are the coefficients specifying the combination that were effectively by the $S_I$-effects ``measured" or prepared for the Higgs in the start of its existence. We are allowed to consider the starting time as a dynamical variable instead of a time because it can be transformed to being essentially the geometrical distance between the decay products $b+\bar{b}$ say. Then it is (essentially) the conjugate variable to the actual mass or energy in the rest frame of the Higgs called here $H_{Higgs}$.

It is not difficult to see that under canonical transformations we can scale $H_{Higgs}$ and $t_{start}$ oppositely by the same factor: 
\begin{align}
	& H_{Higgs}\to \lambda H_{Higgs}, \nonumber \\
	& t_{start}\to \lambda^{-1}t_{start}.
\end{align}
Thus the corresponding transformation of the coefficients $a$ and $b$ is also a scaling in opposite directions 
\begin{align}
	& a\to \lambda^{-1}a, \nonumber \\
	& b\to \lambda b.
\end{align}
We can if we wish normalize to say $ab=1$. The distribution invariant under the canonical transformation will now be a distribution flat in the logarithm of $a$ or of $b$, say 
\begin{align}
	dP\simeq d\log a.
	\label{bphwe52second}
\end{align}
In pracsis it will turn out that the experimentalist has made some extremely crude measurements of both there are some cut offs making it irrelevant that the canonically invariant distribution (\ref{bphwe52second}) is not formally normalizable. With a distribution of this $d\log a$ form it is suggested that very crudely we shall get a \underline{geometrical} average of the results of the two end points possible 1) and 2) above. This means that we in first approximation suggest a resulting replacement for the usual theory Breit-Wigner peak formula being the geometrical mean of the two Breit-Wigners corresponding to the two above discussed extreme case 1) energy prepared: Breit-Wigner with $\Gamma_{SI}+\Gamma_{USUAL}$ and 2) $t_{start}$ prepared: $\Gamma_{USUAL}$ only.

In all circumstances we must normalize the peak in order that the principle of just one future which is even realized in our model is followed.

That is to say that the replacement for the Breit-Wigner in our model becomes crudely 
\begin{align}
	D_{BW\> OUR\> MODEL}(E)=N\sqrt{D_{BW\>\Gamma_{USUAL}+\Gamma_{SI}}(E)
	D_{BW\> \Gamma_{USUAL}}(E)}.
\end{align}

If we assume the $\Gamma_{SI}$ large we may take roughly the broad Breit-Wigner $D_{BW\> \Gamma_{USUAL}+\Gamma_{SI}}(E)$ to be roughly a constant as a function of the actual Higgs rest energy $E$. In this case we get simply 
\begin{align}
	D_{BW\> OUR\> MODEL}(E)\simeq \hat{N}\sqrt{D_{BW\> \Gamma_{USUAL}}(E)}
\end{align}
where $\hat{N}$ is just a new normalization instead of the normalization constant $N$ in foregoing formula.

The crux of the matter is that we argue for that our model modifies the usual Breit-Wigner by taking the \underline{square root} of it, and then normalize it again. The total number of Higgs produced should be (about) the same as usual. But our model predicts a more broad distribution behaving like the square root of the usual one.

\subsection{Please look for this broadening}

This square rooted Breit-Wigner is something it should be highly possible to look for experimentally in any Higgs producing collider and according to the above mentioned bad statistics data from L.E.P. it may already be claimed to have been weakly seen, but of course since the seeing of the Higgs itself were very doubtful at L.E.P. at $115$GeV the broadening is even less statistically supported but it certainly looks promissing.

The tail behaviour of a square rooted Breit-Wigner falls off like 
\begin{align}
	\frac{\text{const.}}{|E-M_H|}
\end{align}
rather than the in usual theory faster fall off 
\begin{align}
	\frac{\text{const.}}{|E-M_H|^2}
\end{align}
where $M_H$ is the Higgs (resonance) mass and $E$ is the actual decay rest system energy.

Let us notice that the integral of the tail in our model (broadened Higgs) leads to a logarithmic dependence 
\begin{align}
	\int \frac{1}{|E-M_H|}dE\simeq \log |E-M_H|.
\end{align}
If for instance we put the $\Gamma_{USUAL}\sim 1$MeV and an effective cut off of $|E-M_H|$ for large values at $\sim 100$GeV and look in a band of $1$GeV size for Higgses, we should find only find 
\begin{align}
	\frac{\log \frac{1\text{GeV}}{1\text{MeV}}}
	{\log \frac{100\text{GeV}}{1\text{MeV}}}
	=\frac{3}{5}=0.6
\end{align}
of the Higgses produced and otherwise visible inside the $1$GeV band. The remaining $0.4$ of them should be further off the central mass $M_H$. This $\frac{1}{|E-M_H|}$ probability distribution should be especially nice to look for because it would so to speak show up at all scales of accuracy of measuring the actual mass of Higgses produced. So there should really be good chances for looking for our broadening as soon as one gets any Higgs data at all.

Let us also remark that the distribution integrating to a logarithm of $|E-M_H|$ obtained by this our second method is in reality very little different from the result obtained by method number one. So we can consider the two methods as checking and supporting each other.

\section{Conclusion}

We have in the present article sought to develop the consequencies of the action $S[path]$ noe being real but having an imaginary part $S_I[path]$ so that $S[path]=S_R[path]+iS_I[path]$ using it in a Fenmann path way integral 
\begin{align}
	\int e^{\frac{i}{\hbar}S[path]}Dpath
\end{align}
understood to be over paths extending over \underline{all} times.

Our first approximation result were that with a bit of optimistic we can make the observable effects of the imaginary part of the action $S_I$ very small in spite of the fact that the supposedly huge factor $\frac{1}{\hbar}$ multiplying $S_I$ in the exponent $e^{\frac{i}{\hbar}(S_R+iS_I)}$ suggests that $S_I$ gives tremendous correction factors in the Feynman path integral. The strong suppression for truly visible effects which we have achieved in the present article seems enough to optimistically say that it is not excluded that there could indeed exist an imaginary part of the action in nature! Since we claimed that it is a less elegant assumption to assume the action real as usual than to allow it to be complex, finding ways to explain that the $S_I$ should not yet have made itself clearly felt would imply that we then presumably \underline{have} an imaginary component $S_I$ of the action!

The main speculations or assumptions arguing for the practical suppression of all the signs of an imaginary action $S_I$ were: 
\begin{enumerate}
\item The classical equations of motion become -at least to a good approximation- given by the variation of the real part $S_R[path]$ of the action alone.
\item While in the classical approximation the imaginary part $S_I$ rather selects which classical solution should be realized.
\item Under the likely assumption that the possibilities to adjust a classical solution to obtain minimal $S_I$ are better by finetuning the solution according to its behavior in the big bang time than today we obtain the prediction that the main simple features of the solution being realized will be features that could be called initial conditions in the sense of concerning a time in the far past. The properties of this solution at time $t$ will be less and less simple -with less and less recognizable simple features- as $t$ increases. This is the second law of thermodynamics being naturally in our model.
\item To suppress the effects of $S_I$ sufficient it is important to have what we above called case 1) meaning that there is one well defined classical path $path_{min}$ with absolute minimal $S_I$ -except for a smaller amount of paths that follow this path $path_{min}$ except for shorter times- being realized. This case 1) is the opposition to case 2) in which there are so many paths with a less negative $S_I$ that they get more likely because of their large number in spite of the probability weight $e^{-\frac{2S_I}{\hbar}}$ being smaller.
\item The contributions to the imaginary action $S_I$ from the relatively short times over which we have proper knowledge -the time of the experiment or the historical times- are so small compared to the huge past and future time spans that the understandable contributions, like the contribution $S_{I\> during\> exp}$ comming under an experiment, drowns and ends up having only small influence on which solution has the absolutely minimal $S_I$. But the huge contributions from far future say we do not understand and in pracsis must consider random (this actually gives us the randomness in quantum mechanics measurements).
\end{enumerate}
In spite of that we talked away the major effects of the imaginary action to determine some phenomenologically not so bad initial conditions mainly, there are still some predictions: 
\renewcommand{\theenumi}{\Alph{enumi}}
\begin{enumerate}
\item For the long times the Universe should go into a state with very low $S_I$ and preferably stay there. Thus the state over long times should be an approximately stable state. Such a prediction of approximate stability fits very well with that presently phenomenological models have a lower bound for the Hamiltonian and being realized after a huge Hubble expansion having brought the temperature to be so low that there is no severe danger for false vacua or other instabilities perhaps accessible if higher energies were accessible in particle collisions.
By the Hubble expansion and the downward approximately bounded Hamiltonian approximately simply vacuum is achieved. Imagining that the vacuum achieved has been chosen via the choice of the solution to have very low (in the sense of very negative) imaginary Lagrangian density $\mathcal{L}_I$ such a situation would just be favourable to reach the minimal $S_I$.
\item In interference experiments where often two for a usually short time separate (roughly) classical solutions or paths are needed for explaining the interference it is impossible to hope that huge $S_I$-contributions from the long future and past time spans can overshadow ($\sim$ dominate out) the imaginary action contribution $S_{I\> during\> exp}$ comming in the interference experiment. The point is namely that from the short time comming $S_{I\> during\> exp}$ can be different for the different interfering paths, but since these paths continue jointly as classical solutions in both past and future they must get exactly the same $S_I$-contributions from the huge past and future time spans. Thus the difference between the $S_{I\> during\> exp}$ for the interfering paths can\underline{not} be dominated out by the longer time spans and their effect must appear to be observed as a disturbance of the interference experiment.
\end{enumerate}
We discussed a lot what is presumably the most promissing case of seeing effects of the imaginary part of the action in an interference experiment: the broadening of the Higgs decay width. In fact an experiment in which a sharp invariant mass measurement of the decay products of a Higgs particle -say Higgs$\to \gamma \gamma$- is performed may be considered a measurement of an interference between paths in which the Higgs particle has ``lived" longer or shorter. Since we suggest that the Higgs contributes rather much to $S_I$ the longer it ``lives" the quantum amplitudes from the paths with a Higgs that live longer may be appreciably more suppressed than those with the Higgs being shorter living. This is what disturbs the interference and broadens the Higgs width.

Our estimates lead to the expectation of crudely a shape of the Higgs peak being more like the square root of the Breit-Wigner form than as in the usual (i.e. $S_I=0$) just a Breit-Wigner.

We hope that this Higgs broadening effect might be observable experimantally. In fact there were if we assume that the Higgs found in Aleph etc at LEP really were a Higgs some excess of Higgs-like events under the lower bound $114$GeV for the mass which could remind of the broadening.

Presumably the here prefered case 1) is the right way -something that in principle might be settled if we knew the whole action form both real and imaginary part- but there is also the possibility of case 2) namely that the realized solution is not exactly the one with the absolutely lowest $S_I$ but rather has a somewhat higher $S_I$ being the most likely due to there being a much higher number of classical solutions with this less extremal $S_I$-value.

While we in the case 1) only may see $S_I$-effects via the interferences in pracsis, we may in case 2) possibly obtain a bias -a correction of the probabilities- due to the $S_I$ even if we for example measure the position of a particle prepared in a momentum eigenstate. Such effects might be easier to have been seen and we thus prefer to hope for -or fit we can say- our model to work in case 1).

It should be stressed that even with case 1) the arguments for the effects of $S_I$ being suppressed are only approximate and practical. So if there were for some reason an exceptional by strong $S_I$ contribution it could not be drowned in the big contributions from past and future but would show up by making the minimal $S_I$ solution be one in which this numerically very big contribution were minimal itself.

This latter possibility could be what one saw when the Super conducting Supercollider (S.S.C.) had the bad luck of not getting funded. It would have produced so many Higgses that it would increased $S_I$ so much that it were basically not possible to find the minimal $S_I$ solution as one with a working S.S.C..

We should at the end mention that we have some other publications with various predictions which were only sporadically alluded to in the present article. For instance we expect the L.H.C. accelerator to be up to similar bad luck as the SSC and we have even proposed a game of letting a random number deciding on restrictions -in luminosity or beam energy- on the running of L.H.C. so that one could in a clean way see if there were indeed an effect of ``bud luck" for such machines.

With mild extra assumptions -that coupling constants may also adjust under the attempt to minimize $S_I$- we argued for a by one of us beloved assumption ``Multiple point principle". This principle says that there are many vacua with -at least approximately- same vacuum energy density (= cosmological constant). Actually we have in an ealier article even argued that our model with such extra assumption even solves the cosmological constant problem by explaining why the cosmological constant being small helps to make $S_I$ minimal. Since the ``Multiple point pronciple"~\cite{rf:6,rf:7} is promissing phenomenologically and of course small cosmological constant is strongly called for it means that the cosmological predictions including the Hubble expansion and the Hamiltonian bottom are quite in a good direction and support our hypothesis of complex action.

\section*{Acknowledgements}
One of us (M.N) acknowledge the Niels Bohr Institute (Copenhagen) for their hospitality extended to him. The work is supported by Grant-in-Aids for Scientific Research on Priority Areas,  Number of Area 763 ``Dynamics of Strings and Fields", from the Ministry of Education of Culture, Sports, Science and Technology, Japan. We also acknowledge discussions with colleagues especially John Renner Hansen about the S.S.C.


\end{document}